\documentclass[pre,twocolumn,epsfig,showpacs,superscriptaddress,floatfix]{revtex4}
\usepackage{graphicx,amsfonts,epsfig}
\usepackage{amsmath}
\usepackage{amssymb}
\usepackage{subfigure}
\usepackage{rotating}

\pacs{\\
81.05.Rm: Porous materials, granular materials, \\
83.10.Rs: Computer simulation of molecular and particle dynamics, \\
61.43.Hv: Fractals, macroscopic aggregates, \\
47.57.J-: Colloidal systems}

\begin{document}

\newcommand{\be}{\begin{equation}}
\newcommand{\ee}{\end{equation}}
\newcommand{\ba}{\begin{aligned}}
\newcommand{\ea}{\end{aligned}}
\newcommand{\bi}{\begin{itemize}}
\newcommand{\ei}{\end{itemize}}
\newcommand{\bn}{\begin{enumerate}}
\newcommand{\en}{\end{enumerate}}
\newcommand{\rv}{\vec{r}}
\newcommand{\vv}{\vec{v}}
\newcommand{\Fv}{\vec{F}}
\newcommand{\im}{\item}
\newcommand{\bc}{\begin{center}}
\newcommand{\ec}{\end{center}}
\newcommand{\ta}{\hat {t}}
\newcommand{\nn}{\hat{n}}
\newcommand{\nij}{\nn _{ij}}
\newcommand{\tij}{\ta _{ij}}
\newcommand {\defeq}{\stackrel{\mbox{\tiny def}}{=}}
\newcommand {\qq}   {\,\qquad}
\newcommand {\qqp}  {\,;\qquad}
\newcommand {\tab} [1] {table\,\ref{tab-#1}}
\newcommand{\ave}[1]{\langle #1 \rangle}
\newcommand{\norm}[1]{\left|\left| #1 \right|\right|}
\newcommand{\abs}[1]{\left| #1 \right|}
\newcommand {\scbox} [1] {\mbox{\scriptsize #1}}
\newcommand{\disty}{\displaystyle}
\newcommand {\as}{\text{I}}
\newcommand {\RR}  { {\mbox{\tiny RR}  } }
\newcommand {\nRR} { {\mbox{\tiny no RR} } }
\newcommand{\degr}{^\circ}
\let\eps = \varepsilon
\newcommand {\Hyp} { {\mbox{$_2$F$_1$}} }
\newcommand{\ww}[1]{\underline{\underline{{\bf #1}}}}
\newcommand{\sti }{\ww{K}}
\newcommand{\stia }{\sti^{(1)}}
\newcommand{\stib }{\sti^{(2)}}
\newcommand{\trs}[1]{\hspace{.05em}\,^{\bf T}\hspace{-.1em}\ww{#1}}
\newcommand{\rig }{\ww{G}}
\newcommand{\rigt }{\trs{G}}
\newcommand{\kk}{{\ww{\mathcal K}}}
\newcommand{\fext}{{\bf F}^{\text{ext}}}
\newcommand{\hhh}{\textsc{h}}
\newcommand{\kkk}{\textsc{k}}
\input epsf

\title{Computer simulation of model cohesive powders:\\
influence of assembling procedure and contact laws on low consolidation states.}

\author{F.~A.~Gilabert}
\email{gilav@us.es}
\affiliation{Faculty of Physics, University of Seville, Avda. Reina Mercedes s/n, 41012 Seville, Spain.}

\author{J.-N.~Roux}
\affiliation{Laboratoire des Mat\'eriaux et des Structures du G\'enie Civil\footnote{
LMSGC is a joint laboratory depending on Laboratoire Central des Ponts et Chauss\'ees,
\'Ecole Nationale des Ponts et Chauss\'ees and Centre National de la Recherche Scientifique.}, Institut Navier,
2 All\'ee Kepler, Cit\'e Descartes, 77420 Champs-sur-Marne, France.}

\author{A.~Castellanos}
\affiliation{Faculty of Physics, University of Seville, Avda. Reina Mercedes s/n, 41012 Seville, Spain.}

\date{October, 2006}

\begin{abstract}

Molecular dynamics simulations are used to
investigate the structure and mechanical properties of a simple
two-dimensional model of a cohesive granular material.
Intergranular forces involve elasticity, Coulomb friction and a
short range attraction akin to the van der Waals force in powders.
The effects of rolling resistance (RR) at intergranular contacts are
also studied. The microstructure of the cohesive packing under low
pressure is shown to depend sensitively on the assembling
procedure which is applied to the initially isolated particles of
a granular gas. While a direct compression produces a final
equilibrated configuration with a similar density to that of
cohesionless systems, the formation of large aggregates prior to
the application of an external pressure results in much looser stable
packings. A crucial state variable is the ratio $P^*= Pa/F_0$ of
applied pressure $P$, acting on grains of diameter $a$, to maximum
tensile contact force $F_0$. At low $P^*$ the force-carrying
structure and force distribution are sensitive to the level of velocity fluctuations in
the early stages of cluster aggregation. The coordination number of packings with RR
approaches 2 in the limit of low initial velocities or large rolling friction.
In general the force network is composed
of hyperstatic clusters, typically comprising 4 to a few tens of
grains, in which forces reach values of the order of $F_0$, joined
by barely rigid arms, where contact forces are very small. Under growing $P^*$, it
quickly rearranges into force chain-like patterns that are more familiar in dense systems.
Density correlations are interpreted in terms of a
fractal structure, up to a characteristic correlation length $\xi$ of the order of
ten particle diameters for the studied solid fractions.
The fractal dimension in systems with RR coincides, within measurement
uncertainties, with the ballistic aggregation
result, in spite of a possibly different connectivity, but is apparently higher without RR.
Possible effects of
micromechanical and assembling process parameters on mechanical strength of packings are evoked.
\end{abstract}

\maketitle

\section{Introduction \label{introduction}}

Granular materials are currently being studied by many research
groups~\cite{HHl98,KI01,HW04,GRMH05}, motivated by fundamental
issues (such as the relations between microstructure and global
properties) as well as by practical needs in civil engineering and
in the food and drug industries. The relation of their
mechanical behavior in quasistatic conditions to the packing
geometry, which depends itself on the assembling procedure,
tends to escape intuition and familiar modelling schemes.

The configuration of the contact networks is hardly accessible to
experiments, even though particle positions are sometimes
measured~\cite{RPBBTB03,XZY04,ASSS04,AsSaSe05}, and some experimental quantitative studies on
intergranular contacts carried out in favorable cases (such as millimeter-sized beads joined by
capillary menisci~\cite{XZY04,KOH04,Fournier05,RiEYRa06}). Intergranular forces are also, most often,
inaccessible to measurements.
Consequently, computer simulation methods of the
``discrete element'' type, as introduced 30 years ago~\cite{CS79},
have proved a valuable tool to investigate the internal states of
granular systems. Simulation methods like molecular dynamics~\cite{HL98}
or ``contact dynamics''~\cite{JEA99,JAM01} have been gaining an increasingly large constituency of users and a wide
range of applications, as witnessed \emph{e.g.,} by recent conference proceedings~\cite{GRMH05}.

Dry assemblies of grains interacting via contact elasticity and
friction, such as sands or glass beads, might form stable packings
of varying solid fraction (typically between 58\% and 64\% for
monosized spheres if they do not crystallize), which deform
plastically in response to changes in stress \emph{direction},
rather than stress intensity. Their elastic or elastoplastic
properties have been studied by discrete simulation (see,
{\emph{e.g.,}~\cite{TH00,SUFL04}), and, in agreement with laboratory
experiments and macroscopic modelling~\cite{DMWood}, found to depend sensitively on the
initial density. Numerical simulation also stressed the importance
of additional variables such as coordination number~\cite{AgRo05}
and fabric~\cite{BR90,RR04}, and it has often been applied to the
study of quasi-static stress-strain behavior of granular
assemblies (refs.~\cite{TH00,RC02,RR04,SUFL04} are a few examples among a
large literature).

\emph{Cohesive} grains exhibit much larger variations in their
equilibrium densities, and they are sensitive to stress
\emph{intensity} as well as direction~: on increasing the
confining pressure, the specific volume of a clay can irreversibly
decrease by a factor of 4~\cite{MIT93}. Likewise, series of
experiments carried out in the Seville group on model
powders~\cite{C05} (xerographic materials) in which the strength
of van der Waals attraction is controlled by additives covering
part of the grain surfaces, reveal a similar variation of porosity
with confining pressure. It is notable that such packings of
particles of rotund shape and nearly the same size can stay in
mechanical equilibrium at much lower solid fractions (down to
25-30\%) than cohesionless granular systems.

Despite this wider
variety of equilibrium structures and mechanical behaviors,
cohesive granular materials have much less frequently been investigated by
numerical simulation than cohesionless ones.

Some of the recent numerical studies, such as those of
refs.~\cite{YZY00,YZY03a} have investigated the packing structures
of spherical beads deposited under gravity, depending on
micromechanical parameters, including adhesion strength. Another
set of publications report on simulations of the dynamical
collapse and compaction, both in
two~\cite{KBBW02,KBBW03,WoUnKaBr05} and three~\cite{GUKWK05}
dimensions, the main results being the relations between density
and pressure increments, and their dependence on micromechanical
parameters. Some works focussed on the fracture of bound particle
assemblies in static~\cite{DeYB02,DeYB04} or dynamic~\cite{ThLi04}
conditions, others on wet bead packs in which cohesion stems from
liquid bridges joining neighboring particles, investigating the
structure of poured samples~\cite{YZY03b} or the shear
strength~\cite{RiEYRa06} of such materials. These two latter types
of studies deal with relatively dense materials, as does the
numerical biaxial compression test of~\cite{LU05}. Flow of
cohesive materials has also been addressed in recent
publications~\cite{RoRoCh05,RoRoWoNaCh06,BrGrLaLe05}.

Yet, numerical studies of the \emph{mechanics} of loose,
solid-like cohesive granulates are quite scarce. This contrasts
with the abundant literature on the \emph{geometry} of model loose
particle packings and colloidal aggregates, which tend to form
fractal structures. Refs.~\cite{SMI90,MEA99} are useful overviews
of aggregation processes and the geometric properties of the
resulting clusters, as obtained by numerical simulation. In such
processes, particle aggregates are usually regarded as
irreversibly bound, rigid solids, while the interaction between
separate clusters reduces to a ``sticking rule'', so that both
intra- and inter-aggregate mechanical modelling is bypassed.
Interestingly, one simulation study~\cite{BRH02} shows that
structures resulting from geometric deposition algorithms are not
always stable once a mechanical model is introduced.

It seems that numerical simulations of both geometric and
mechanical properties of loose granular assemblies forming solid
aggregates are still lacking.

The present paper addresses part of this issue. It reports on
numerical simulation studies of cohesive granular materials, with
the following specificities: \bi \im the \emph{assembling process}
is simulated with the same mechanical model as applied to
solid-like configurations, and its influence on the packing
microstructure is assessed ; \im special attention is paid to
\emph{loose} particle packings \emph{in equilibrium} under
vanishing or low applied pressure; \im \emph{both} geometric and
mechanical properties are investigated ; \im isotropic and
homogeneous systems are studied, as representative samples for
bulk material properties. \ei

We consider a simple model system in two dimensions,
introduced in section~\ref{sec:model}, along with the numerical
simulation procedure. Despite its simplicity we shall see that this model
yields results that are amenable to comparisons with experimental situations.

Section.~\ref{sec:preparation} is devoted to the important issue of
the procedure to prepare samples, and its influence, as well as that of
micromechanical features such as rolling resistance (RR), on final
density and coordination number in solid packings in equilibrium.
In Section~\ref{sec:mech} we investigate the force
distributions and force patterns of the equilibrated loose
configurations under vanishing or low applied pressure. Some
specific aspects of the force-carrying structures in low density
assemblies will be studied and related to the assembling process.
In Section~\ref{sec:geometry}, we characterize the geometry and
density correlations in loose samples, resorting to the fractal
model traditionally employed for colloidal aggregates.
Finally we conclude in Section~\ref{sec:final} with a few remarks
about future improvements and further developments of this work,
some of which will be presented in a forthcoming
publication~\cite{paper2}.

\section{Model material\label{sec:model}}

\subsection{System definition, equations of motion\label{subsec:system}}
We consider a two-dimensional model material: an assembly of $N$
disks with diameters $(d_i)_{1\le i\le N}$ uniformly distributed
between $a/2$ and $a$. The maximum diameter, $a$, will be used as
unit of length. The mass of grain $i$ is $m_i= d_i^2/a^2$ and its
moment of inertia $I_i= m_i d_i^2/8$, \emph{i.e.} disks are
regarded as homogeneous bodies and the mass of a disk of maximum
diameter $a$ is the unit of mass.

The disks are enclosed in a rectangular cell the edges of which
are parallel to the axes of coordinates $x_1$ and $x_2$, with
respective lengths $L_1$ and $L_2$. Periodic boundary conditions
are used, thereby avoiding wall effects. Neighboring grains, say
$i$ and $j$, might interact if they are brought into contact or
very close to each other, hence a force $\Fv _{ij}$ and a moment
${\Gamma}_{ij}$ exerted by $i$ onto $j$ at the \emph{contact
point}. Simulations do not model material deformation in a contact
region, but consider overlapping particles, and the contact point
is defined as the center of the intersecting surface of the two
disks. In the case of an interaction without contact, the force
will be normal to the surfaces at the points of nearest approach,
and therefore carried by the line of centers. Let $\rv _i$ denote
the position of the center of disk $i$. $\rv _{ij}=\rv _j -\rv _i$
is the vector joining the centers of $i$ and $j$, and $h_{ij} =
\vert \rv _{ij} \vert - (d_i+d_j)/2$ their overlap distance. The
degrees of freedom, in addition to the positions $\rv _i$, are the
angles of rotation $\theta _i$, velocities $\vv_i$, angular
velocities $\omega _ i= \dot \theta _ i$ of the grains ($1\le i\le
N$), the dimensions $(L_\alpha)_{\alpha=1,2}$ of the cell
containing the grains and their time derivatives, through the
strain rates:
$$\dot \epsilon_{\alpha} = -\dot L_{\alpha}/L_{\alpha}^0,$$
in which $L_{\alpha}^0$ denotes the initial size for the
corresponding compression process. The time evolution of those
degrees of freedom is governed by the following equations.
\be
   m_i \frac{d^{2}\vec{r}_i}{d t^{2}} =
   \sum_{j=1}^N
    \Fv_{ij}\label{eqn:newton} \ee
\be
   I_i \frac{d\omega_i}{dt} =\sum_{j=1}^N \Gamma_{ij}\label{eqn:moment}
\ee
\be
\ba
    M \frac{d^2\epsilon _\alpha }{dt^2} &=
        \sigma^I_{\alpha\alpha} -
        \sigma^M_{\alpha\alpha}\\
    \sigma^M_{\alpha\alpha} &=
        \frac{1}{A} \sum_{i=1}^N
        \left[
            m_i v_{i,\alpha} ^2
            +
            \sum_{j\ne i}  F^{(\alpha)}_{ij} r^{(\alpha)}_{ij}
        \right]
\ea
\label{eqn:cell}
\ee
In Eqns.~\eqref{eqn:newton} and \eqref{eqn:moment}, only those
disks $j$ interacting with $i$, \emph{i. e.} in contact or very
close, will contribute to the sums on the right-hand side. In
Eqn.~\eqref{eqn:cell}, $\sigma^I_{\alpha\alpha}$ is the externally
imposed stress component, $\sigma^M_{\alpha\alpha}$ is the
measured stress component, resulting from ballistic momentum
transport and from the set of intergranular forces $\vec{F}_{ij}$,
$A=L_1 L_2$ denotes the cell surface area, and $M$ is a
generalized inertia parameter.

Stresses $\sigma_{11}$ and $\sigma_{22}$, rather than strains or
cell dimensions, are controlled in our simulation procedure. Note
that compressions are counted positively for both stresses and
strains. Eqn.~\eqref{eqn:cell} entails that the sample will expand
(respectively, shrink) along direction $\alpha$ if the
corresponding stress $\sigma^M_{\alpha \alpha}$ is larger (resp.,
smaller) than the requested value $\sigma^I_{\alpha\alpha}$, which
should be reached once the system equilibrates. This
\emph{barostatic} method is adapted from the ones initially
proposed by Parrinello and Rahman \cite{PARA80,PARA81,PARA82} for
Hamiltonian, molecular systems.

The choice of the ``generalized mass'' $M$ is rather arbitrary,
yet innocuous provided calculations are restricted to small strain
rates. In practice we strive to approach mechanical equilibrium
states with good accuracy, and choose $M$ in order to achieve this
goal within affordable computation times. We usually attribute to
$M$ a value equal to a fraction of the sum of grain masses (3/10
in most calculations), divided by a linear size $L$ of the cell.
This choice is dimensionally correct and corresponds to the
appropriate time scale for strain fluctuations in the case of a
thermodynamic system.

\subsection{Interaction law \label{sec:forces}}

The \emph{contact law} in a granular material is the relationship
between the relative motion of two contacting grains and the
contact force. As we deal with particles that may attract one
another at short distance without touching, the law governing
intergranular forces and moments is best referred to simply as the
\emph{interaction law}.

Although the interaction we adopted is based on the classical linear ``spring-dashpot" model
with Coulomb friction
for contact elasticity, viscous dissipation and sliding,
as used in many discrete simulations of granular media~\cite{HL98,SEGHL02,BrGrLaLe05,RoRoWoNaCh06},
some of its features (short-range attraction and rolling resistance) are
less common ; moreover, one can think of different implementations of
the Coulomb condition, depending on which parts of normal and tangential
force components are taken into account. Therefore, for the sake of clarity and completeness,
we give a full, self-contained presentation of the interaction law below.

We express intergranular forces in a mobile system of coordinates
with axes oriented along the normal unit vector $\nn _{ij}$ (along
$\rv _{ij}$) and the tangential unit vector $\ta _{ij}$ ($\nn
_{ij},\ta _{ij}$ is a direct base in the plane), and use the convention that
repulsive forces are positive.

The intergranular force $\vec{F}_{ij}$, exerted by grain $i$ onto
its neighbor is split into its normal and tangential components,
$\vec{F}_{ij}=N_{ij}\nn _{ij}+T_{ij}\ta _{ij}$ thus defining
scalars $N_{ij}$ and  $T_{ij}$. $N_{ij}$ comprises a static term
 depending on the distance between disk centers, combining contact elasticity and distant,
van der Waals type attraction, as shown on Fig.~\ref{fig:force}a,
and a velocity-dependent viscous term $N^v_{ij}$. $T_{ij}$
(Fig.~\ref{fig:force}b) is due to the tangential elasticity in the
contact, and is limited by the Coulomb condition.
\begin{figure*}[!hbt]
   \begin{tabular}{cc}
       \psfig{file=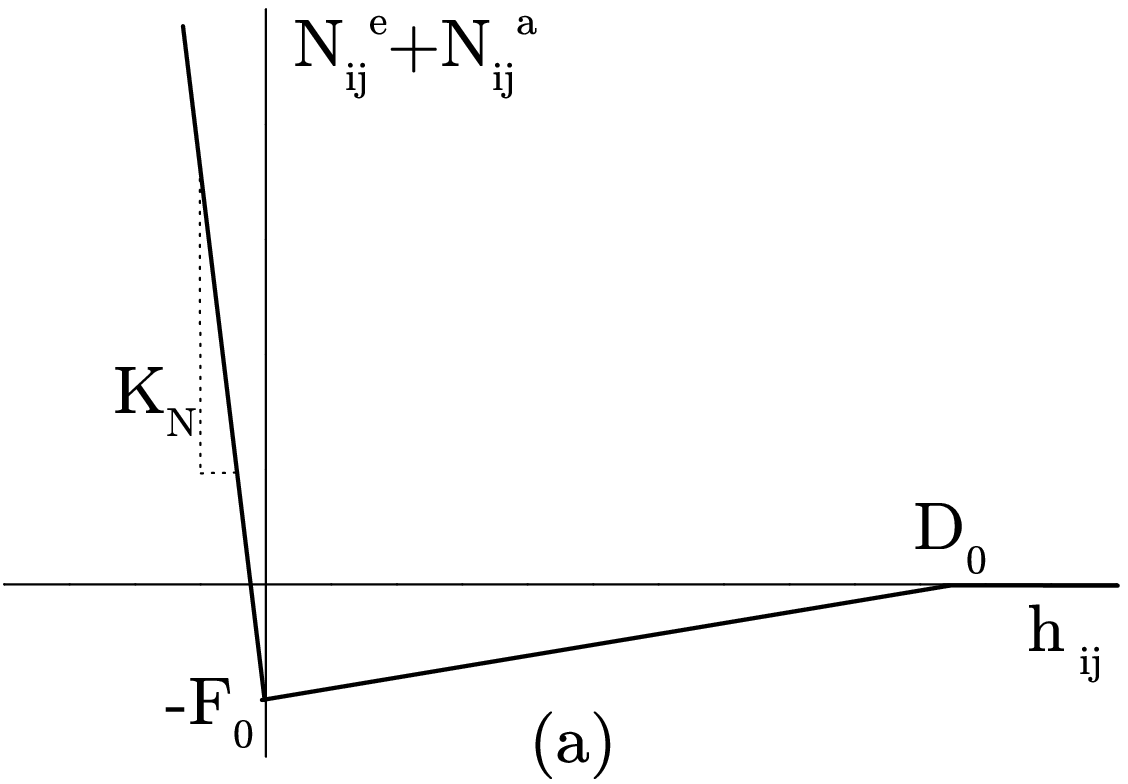,height=5.5cm,angle=0} &
       \psfig{file=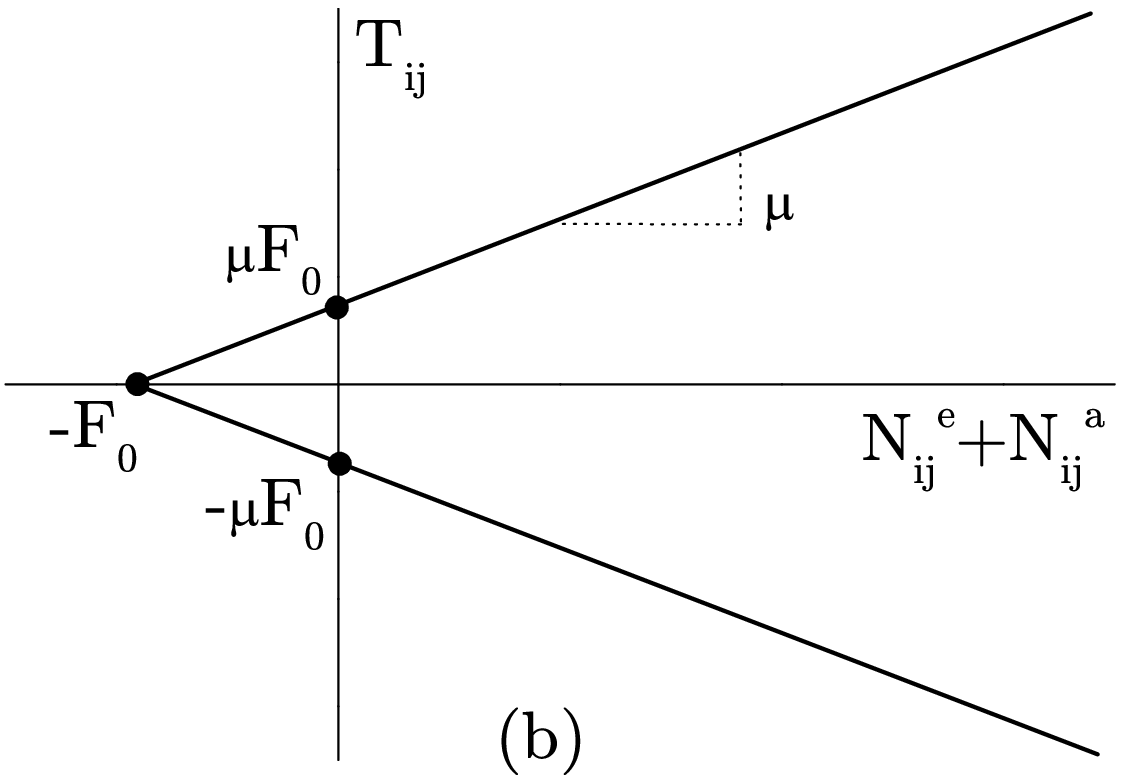,height=5.5cm,angle=0}
   \end{tabular}
   \caption{Graphical representation of the model for the adhesive elastic contact force
    as a function of the distance between the surfaces of particles $i$ and $j$, $h_{ij}$.
    (a) The elastic normal force consists of a repulsive Hookean part $N_{ij}^e$ plus
    a linearized attractive part $N_{ij}^a$. (b) The elastic tangential force is limited by
    the Coulomb cone (adhesion shifting its tip to $-F_0$ on the normal force axis).}
   \label{fig:force}
\end{figure*}
If disks $i$ and $j$ are not in contact, both the tangential component of force
$\vec{F}_{ij}$ and the viscous part of the normal component vanish, while $i$ and $j$
still attract each other if the gap ($h_{ij}\ge 0$)
between their surfaces is smaller than the \emph{attraction range} $D_0$ ($0 \le h_{ij} \le D_0$):
\be
\vec{F}_{ij}=N_{ij}^{a} \nn _{ij}\ \text{with }\ N_{ij}^{a} = -F_0~(1 - \frac{h_{ij}}{D_0})~\nn_{ij}
\label{eqn:attrac}
\ee
This expression is a linear approximation of a realistic van der
Waals force law (see Fig.~\ref{fig:force}a), and contains two
essential parameters, maximum attractive force $F_0$, range, $D_0$. Typically,
$F_0$ is of the order of $\gamma l$, $\gamma$ being a superficial
energy, $l$ the typical size of asperities \cite{Forsyth00} and
$D_0$ is in the nanometer range.

In the case of contacting disks ($h_{ij}<0$), the attractive term
$N_{ij}^{a}$ is kept constant, equal to $-F_0$, while strains in
the contact region result in normal ($N_{ij}^e$) and tangential
($T_{ij}$) elastic forces. It is also assumed that a viscous normal
term $N^v_{ij}$ opposes relative normal displacements. One thus
writes:
\begin{equation}
    \Fv _{ij} = ( N_{ij}^e + N_{ij}^{v} - F_0 )~\nn_{ij} + T_{ij}~\ta_{ij}
\label{eqn:totalforce}
\end{equation}
The different terms introduced in Eqn.~\eqref{eqn:totalforce} are
defined according to the following models. First, $$N_{ij}^e =-K_N
h_{ij}$$ is the linear elastic unilateral repulsion, due to the
normal deflection $-h_{ij}$ in the contact as the disks are
pressed against each other. $K_N$ is the normal stiffness
coefficient, related to the elastic moduli of the material the
grains are made of.

The viscous normal force opposes the normal relative receding
velocity $\delta v_{ij}^N = \nn_{ij}\cdot(\vv_{j}-\vv_{i})$ as
long as the contact persists. The relative normal motion of two
disks $i$ and $j$ in contact is that of an oscillator with viscous
damping, and $\eta_{ij}$ is the damping coefficient. We choose its
value as a constant fraction $\zeta$ of the critical damping
coefficient, \be \eta_{ij} = \zeta \sqrt{\frac{4 K_N m_i m_j}{m_i
+ m_j}}. \label{eqn:damping} \ee This is equivalent to the choice
of a constant restitution coefficient in normal collisions if
$F_0=0$. In the presence of attractive forces the apparent
restitution coefficient in a collision will depend on the initial
relative velocity, and will be equal to zero for small values,
when the receding velocity after the collision will not be able to
overcome the attraction and separate the particles. The minimum
receding velocity for two particles of unit mass (\emph{i.e.}, of
maximum diameter $a$) to separate is $V^*\sqrt{2}$, with \be
V^*=\sqrt{F_0D_0}. \label{eqn:vsep} \ee

The elastic tangential force in contact $i,j$ is linearly related
to the \emph{elastic part} $\delta u _{ij}^T$  of the total
relative tangential displacement $\Delta u^T _{ij}$, as
$$T_{ij} = K_T~\delta u_{ij}^T,$$ and
is subject to the Coulomb inequality. $K_T$ is the tangential stiffness coefficient.
$\Delta u^T _{ij}$ can be updated for all
closed contacts according to
$$\frac{d\Delta u _{ij}^T}{dt} = (\vv_{ij} \cdot \ta _{ij})$$
and vanishes as soon as
the contact opens. Its elastic part satisfies
$$\frac{d\delta u_{ij}^T}{dt} =  H\left(\frac{\mu N_{ij}^e}{K_T}
- \vert \delta u^T_{ij} \vert \right) ~ (\vv_{ij} \cdot \ta _{ij})$$
in which $H$ denotes the Heaviside function. This last equation
introduces the \emph{friction coefficient} $\mu$. It is important
to note that the Coulomb inequality, \be \vert T_{ij}\vert \le \mu
N^e_{ij}, \label{eqn:coulomb} \ee applies to the sole repulsive
elastic component of the normal force (see Fig.~\ref{fig:force}b).
We chose not to implement any tangential viscous force.

The moment that disk $i$ exerts onto its contacting neighbor $j$,
of radius $R_j$, in its center, is denoted by $\Gamma_{ij}$ in
Eqn.~\eqref{eqn:moment}. It is first due to the tangential contact
force, then to a possible moment $\Gamma_{ij}^{~r}$ of the force
density distribution within the contact region. One thus writes:
\be \label{eqn:gamma}
    \Gamma_{ij} = -T_{ij} R_j + \Gamma_{ij}^{~r}.
\ee
$\Gamma_{ij}^{~r}$ is most often neglected on dealing with smooth,
convex particle shapes, because the contact region is very small
on the scale of the particle radius.

To model rolling resistance (RR), like in \cite{TOST02}, we
introduce a rotational stiffness parameter $K_r$ and a rotational
friction parameter $\mu_r$ in contacts, so that rolling elasticity
and rolling friction are modelled just like sliding elasticity and
friction. One thus writes
$$\Gamma_{ij}^{~r} = K_r~\delta\theta_{ij},$$
while enforcing the inequality
\be
K_r |\delta\theta_{ij}| \le
\mu_r N_{ij}^e. \label{eqn:ineqRR}
\ee
This involves the
definition of $\delta\theta_{ij}$ as the elastic part of the total
relative rotation $\Delta\theta_{ij}$. The total relative rotation
angle satisfies
$$\frac{d\Delta \theta _{ij}}{dt} = \omega_j - \omega_i,$$
while the equation for $\delta\theta_{ij}$ is
$$\frac{d\delta \theta_{ij}}{dt} =  H\left(\frac{\mu _r N_{ij}^e}{K_r} -
\vert \delta \theta_{ij} \vert \right) ~ (\omega _j - \omega _i).$$
Parameters $K_r$ and $\mu _r$ are often related to the size of a
contact region \cite{KBBW02}. $K_r$ is dimensionally the product
of a stiffness by the square of a length, which is of the order of
the contact size. In the following we set $K_r$ to
$10^{-4}a^2K_N$, while $\mu _r$, which has the dimension of a
length, is chosen equal to $10^{-2}\mu a$.
The motivation for the introduction of RR into our model is twofold. First, cohesive
particles are usually small (typically less than $30\mu$m in size)
and irregular in shape. Contacts between grains are likely to
involve several asperities, and hence some lateral extension, of
the order of the distance between asperities, however small the
normal deflection $-h$.
Then, it will be observed that even quite a small rotational friction has a notable
influence on the microstructure of cohesive packings.
%

\subsection{Control parameters and dimensional analysis \label{subsec:parameters}}

In this section we present the dimensionless parameters which
express the relative importance of different physical phenomena.
Such parameters enable qualitative comparisons with real
materials, bearing in mind that the present model is admittedly an
idealization of real powders and that our simulations do not aim
at quantitative accuracy.

Dimensionless numbers related to contact behavior are the
\emph{reduced interaction range} $D_0/a$, the friction coefficient $\mu$, the
viscous damping parameter $\zeta$, and the stiffness parameter $\kappa$.

Under the attractive force $-F_0$, the elastic deflection of
one contact is
\be \label{eqn:h0}
    h_0=F_0/K_N
\ee
The \emph{stiffness parameter} $\kappa \defeq aK_N / F_0$ characterizes the
amount of elastic deflection $h_0$ under contact force $F_0$,
relative to grain size $a$ ($h_0/a = \kappa^{-1}$). A suitable analogous definition for
Hertzian spheres in three dimensions would be $\kappa=(Ea^2/F_0)^{2/3}$.

The dimensionless number $h_0/D_0$ is the ratio of elastic to
adhesive stiffnesses, and its physical meaning is similar to that
of the Tabor parameter $\lambda = (1/D_0)(\gamma ^2 a
/E^2)^{1/3}$~\cite{MAU00} for a Hertzian contact between spheres
of diameter $a$ when the material Young modulus is $E$ and the
interfacial energy is $\gamma$ (more precisely, the equilibrium
normal deflection $h_0$, due to adhesion, in the contact between
an isolated pair of grains, satisfies $\lambda \sim (h_0/D_0)
^{1/3}$ in this case).

The viscous damping parameter, $\zeta$, corresponds to a normal
restitution coefficient $e_N =\exp [-\pi \zeta /\sqrt{1-\zeta
^2}]$ in the absence of cohesion ($F_0=0$).

In our calculations we set $\zeta=0.8$, corresponding to a high
viscous dissipation in collisions, or a very low restitution
coefficient in binary collisions. Models with a constant $\zeta$
were adopted in other published simulation
works~\cite{SEGHL02,SRSvHvS05}, although little is known about
dissipation in collisions. $\zeta$ is known to influence the
packing structures obtained in the initial assembling
stage~\cite{ZLY01,SEGHL02}, but we did not investigate its effects
in the present study. The simulations reported
in~\cite{YZY00,YZY03a} use the viscous force model introduced in
ref.~\cite{BrSpHePo96}, with a choice of parameters corresponding
to strongly overdamped dynamics (\emph{i.e.}, analogous to $\zeta
\gg 1$ in our case).

In addition to those control parameters determined by the contact
behavior, other dimensionless numbers are introduced by the
loading or the process being applied to the material. The effect
of the external pressure, compared to the adhesion strength, is
characterized by a dimensionless \emph{reduced pressure} $P^*$:
\be
 P^* \defeq P a / F_0.
 \label{eqn:defP}
\ee
In the present paper, we focus on the assembling process and the
low $P^*$ range. As we shall see below
(Sec.~\ref{sec:preparation}) low density, tenuous structures are
then stabilized by adhesion, and the relevant force scale is
$F_0$. However, as briefly reported in~\cite{GRC05}, such
structures tend to collapse upon increasing $P^*$. These phenomena
will be the subject of another paper~\cite{paper2}. Wolf
\emph{et al.}~\cite{WoUnKaBr05} introduced a dimensionless stress
proportional to $P^*$, and observed, in numerical simulations, stepwise increases in pressure
to produce large dynamical collapse effects around $P^*=1$. The importance of $P^*$ was
also stressed in simulations of cohesive granular flow, in which
the effects of cohesion on rheological laws were expressed in
terms of a cohesion number defined as $1/P^*$~\cite{RoRoWoNaCh06}.
In three dimensions, $P^*$ should be defined as $a^2P/F_0$.

For large reduced pressures, externally imposed forces dominate
the adhesion strength, and one should observe behaviors similar to
those of confined cohesionless granular materials. For $P^* > 1$,
the relevant force scale is $aP$. The influence of $\kappa$, which
should then be defined as $\kappa \defeq K_N / P$, so that the
typical contact deflection $h$ satisfies $h/a \propto
\kappa^{-1}$, was studied in simulations of grains without
adhesion~\cite{CR03}. Whatever the reference force used to define
it, the limit of rigid grains is $\kappa \to +\infty$. With
relatively soft grains (say, $\kappa$ below $10^3$), a significant
number of additional contacts appear in dense configurations, due
to the closing of gaps between near neighbors. Such a $\kappa$
parameter defined with reference to pressure, in the case of
contacts ruled by Hertz's law between spherical grains made of a
material with Young modulus $E$, should be chosen as
$\kappa=(E/P)^{2/3}$, in order to maintain $h/a \sim \kappa^{-1}$.

In order to stay within the limit of rigid
grains both for small and large $P^*$, we choose quite a large value
of $\kappa=K_Na/F_0 $:  $\kappa =10^4$ or $\kappa =10^5$.

Table~\ref{tab:param} summarizes the values (or the range of
values) of dimensionless parameters in the simulations presented
below.
\begin{table}[!htb]
\centering
\begin{tabular}{|c|c|c|c|c|c|c|c|}\cline{1-8}
$\mu$  & $\zeta$ &  {\Huge \strut} $\kappa$& ${\disty \frac{K_T}{K_N}\strut }$ & ${\disty \frac{D_0}{a}}$ &
${\disty \frac{K_r}{K_Na^2}}$& ${\disty \frac{\mu_r}{a}}$& $P^*$   \\ \hline
$0.15$, $0.5$&$0.8$&$10^5$, $10^4$&$1$&$10^{-3}$&$10^{-4}$&0, $10^{-2}\mu $&$0$, $0.01$\\ \hline
\end{tabular}
\caption{\label{tab:param} Values of dimensionless model
parameters used in most simulations. Note that $h_0/D_0$ is fixed
by ${\disty \kappa=\frac{K_N a}{F_0}}$ and $D_0/a$ to $10^{-2}$ or $10^{-1}$. In the absence
of cohesion, or for values of $P \ge F_0/a$, $\kappa$ is defined
as $K_N/P$}
\end{table}
In addition to those values of the parameters, adopted as a
plausible choice for realistic orders of magnitudes, some
calculations were also performed with deliberately extreme
choices, such as very large RR ($\mu_r=0.5a$) or absence of friction
($\mu=0$ and $\mu_r=0$), in order to better explore some connections
between micromechanics and macroscopic properties. The
corresponding results will be described in
Section~\ref{sec:mech}.

The definition of dimensionless parameters, suitably generalized
to three-dimensional situations as $P^*=a^2 P/ F_0$ and
$\kappa\simeq (Ea^2/F_0)^{2/3}$ for spherical particles of
diameter $a$, enables one to discuss qualitative features and
orders of magnitude in the model system defined with the
parameters of Table~\ref{tab:param} with comparisons to some
cohesive packings studied in the laboratory.

When adhesive forces are due to liquid menisci joining neighboring
particles, we should take $F_0 \sim \gamma a$, where $\gamma$ is
the surface tension. $P^*=1$ corresponds then to confining
pressure $P$ in the range of 10-100~Pa for millimeter-sized
particles, taking standard values for $\gamma$. Those are rather
low pressures in practice, which are comparable, \emph{e.g.}, to
the ones caused by the weight of a typical laboratory sand sample.
Thus wet granular materials are commonly under reduced pressures
$P^*$ of order 1 or larger, and are not observed with much lower
solid fractions than dry
ones~\cite{XZY04,KOH04,Fournier05,RiEYRa06}.

The cohesive powders studied in
refs.~\cite{WVC01,MiguelAngel,CVQ05,C05} are xerographic toners with
typical particle diameter  $a \sim10$ $\mu$m. $F_0$, the van der
Waals attractive force, is a few tens of nN, and the range $D_0$ is
several nanometers \cite{KRU67}. Therefore, a reduced pressure
$P^*=0.01$ would correspond to about $1$~Pa in the experimental
situation \cite{VQC04}. This is an initial state of very low
consolidation stress, which is present in a powder under gravity,
provided a controlled gas flow, going upwards through the powder,
counterbalances part of its weight~\cite{VQC04}. As to contact
stiffnesses, our values of $h_0/a$ would correspond to $E \sim
0.1$~GPa (for $K_N=10^4F_0/a$) or $3.2$~GPa (for $K_N=10^5F_0/a$),
while the ratio $D_0/a$ would imply an interaction range of 10~nm.
This gives us the correct orders of magnitudes for the toner
particles, those being made of a relatively soft solid (polymer,
such as polystyrene) with $E \sim 3-6$~GPa. Xerographic toner
particles appear to undergo plastic deformation in the
contacts~\cite{MiguelAngel,MAPO84,QCV01,GKC06}. Plastic deflections
of contacts are accounted for in the model of ref. \cite{LU05},
applied to the simulation of a biaxial compression of a dense
powder. In our study, for simplicity's sake, and because we expect
macroscopic plasticity of loose samples to be essentially related to
the collapse of tenuous structures, we ignored this feature.
\subsection{Equilibrated states \label{subsec:equilibrium}}
Although numerical simulations of the quasistatic response of
granular materials requires by definition that configurations of
mechanical equilibrium should be reached, equilibrium criteria are
sometimes left unspecified, or quite vaguely stated in the
literature. Yet, in order to report results on important, often
studied quantities like the coordination number or the force
distribution, it is essential to know which pairs of grains are in
contact and which are not. Due to the frequent occurrence of small
contact force values, this requires forces to balance with
sufficient accuracy. We found that the following criteria allowed
us to identify the force-carrying structure clearly enough. We use
the typical intergranular force value $F_1 = \text{max}(F_0,Pa)$
to set the tolerance levels. A configuration is deemed
\emph{equilibrated} when the following conditions are fulfilled:
\bi \im the net force on each disk is less than $10^{-4}F_1$, and
the total moment is lower than $10^{-4}F_1a$; \im the difference
between imposed and measured pressure is less than $10^{-4}
F_1/a$; \im the kinetic energy per grain is less than $5 \cdot
10^{-8}F_1a$. \ei We observed that once samples were equilibrated
according to those criteria, then the Coulomb
criterion~\eqref{eqn:coulomb}, as well as the rolling friction
condition~\eqref{eqn:ineqRR} were satisfied as \emph{strict}
inequalities in all contacts. No contact is ready to yield in
sliding, and with RR no contact is ready to yield in rolling
either.
\section{Assembling procedure \label{sec:preparation}}
It has been noted in experiments~\cite{MIT93} and
simulations~\cite{TH00,AgRo05,SEGHL02} that the internal structure
and resulting behavior of solid-like granular materials is
sensitive to the sample preparation procedure, even in the
cohesionless case.

In the case of powders, it has been observed that the
sedimentation in dry nitrogen (to minimize the capillary effects
of the humidity on the interparticle adhesion) of a previously
fluidized bed produced reproducible states of low solid fractions
(down to $10-15\%$) \cite{VCRPMW00,CVQ01}. This initial state
under such a low consolidation, as we commented in
\ref{subsec:parameters}, plays a decisive role on the evolution of
the dynamics of powder packing. That is, appreciable differences
in initial states will lead to considerable ones in final packings
\cite{CVQ05}. This is mainly due to the role of aggregation, which
we shall analyze in the second part of this section.

The motivation of this section is to investigate the dependence
on packing procedure in a cohesive granular system, the first step
being to obtain stable equilibrated configurations with low
densities. For comparison, some simulation results are presented
for the same model material with no cohesion.

Specimens were prepared in two different ways, respectively
denoted as \emph{method 1} and \emph{method 2}, and the resulting
states are classified as \emph{type 1} or \emph{type 2}
configurations accordingly.

Due to our choice of boundary conditions, our samples will be
completely homogeneous, under a uniform (isotropic) state of
stress. This choice is justified by the complexity of seemingly
more ``realistic" processes, such as gravity deposition, due to
the influence of many material (such as viscous dissipation, as
recalled in Section~\ref{subsec:parameters}) and process
parameters. Both pouring rate and height of free fall should be
kept constant during such a \emph{pluviation} process in order to
obtain a homogeneous packing~\cite{ZLY01,ERCCD05} with
cohesionless grains. Cohesive ones, because of the irreversible
compaction they undergo on increasing the pressure, would end up
with a density increasing with depth. Hence our choice to ignore
gravity in our simulations. Our final configurations should be
regarded as representative of the \emph{local} state of a larger
system, corresponding to a local value of the confining stress.
\subsection{Method 1 \label{sec:method1}}
In simulations of cohesionless granular materials, a common
procedure~\cite{TH00,Makse04,SUFL04} to prepare solid samples
consists in compressing an initially loose configuration (a
``granular gas"), without intergranular contacts, until a state of
mechanical equilibrium is reached in which interparticle forces
balance the external pressure (further compaction being prevented
by the jamming of the particle assembly). We first adopted this
traditional method, hereafter referred to as method 1, to assemble
cohesive particles.

In this procedure, disks are initially placed in random
non-overlapping positions in the cell, with zero velocity. We
denote such an initial situation as the $I$-state. Then the
external pressure is applied, causing the cell to shrink
homogeneously. Thus contacts gradually appear and the
configuration rearranges until the system equilibrates at a higher
density.

Examples of equilibrated configurations are shown on
Fig.~\ref{fig:confmethod1}, with and without cohesion.
\begin{figure*}[!htb]
\centering
\subfigure[No cohesion, $K_N/P=10^5$]
{
  \resizebox{0.98\columnwidth}{!}{\includegraphics*{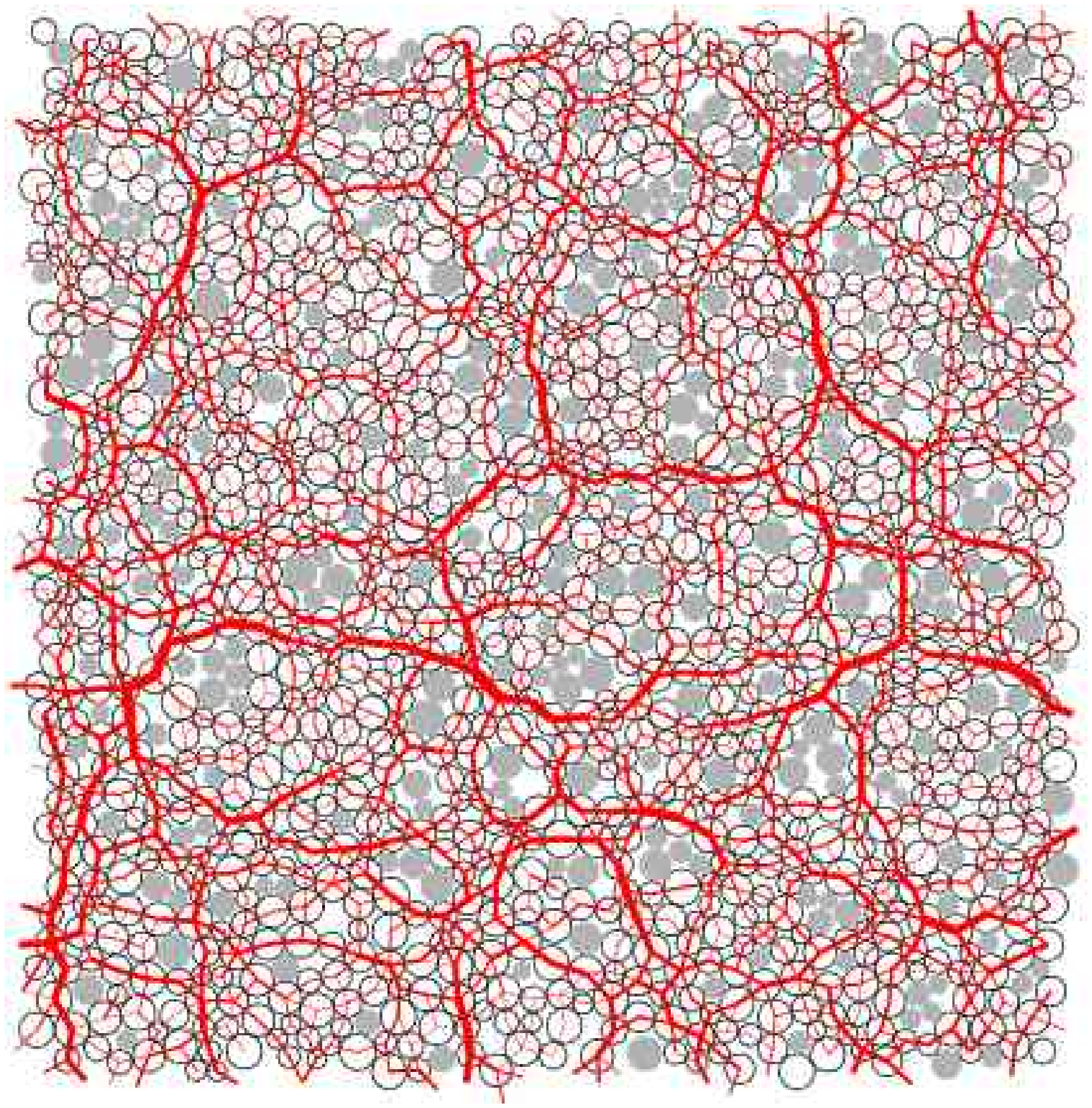}}
  \label{fig:nadm1}
}
\subfigure[Cohesive system, $\kappa=10^5$, $P^*=0.01$]
{
  \resizebox{0.98\columnwidth}{!}{\includegraphics*{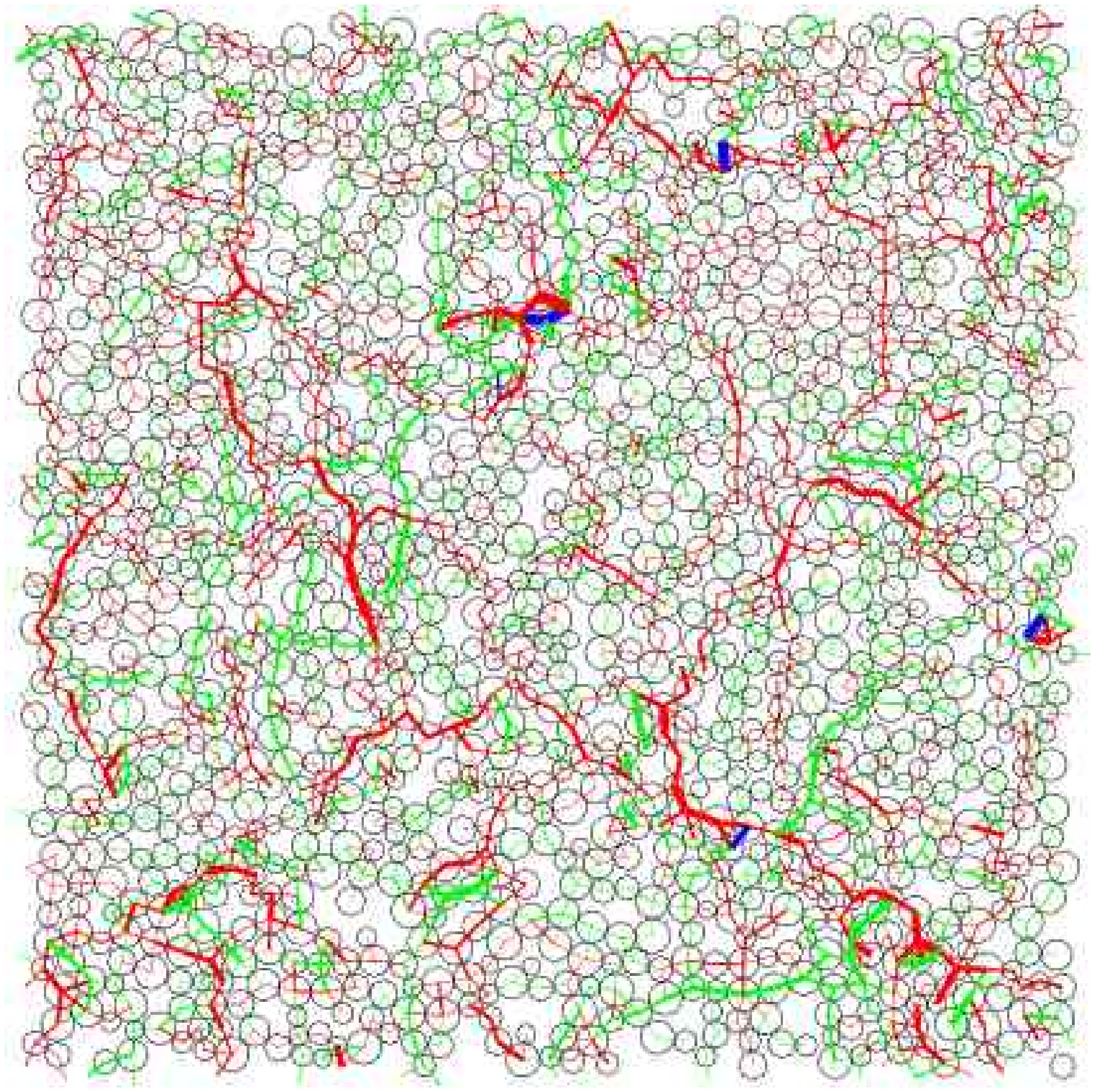}}
  \label{fig:adm1}
}
\caption{(Color online) Aspect of force-carrying
structures in cohesionless and cohesive
samples. Contact forces are displayed with the usual convention
that the width of the lines joining the centers of interacting
pairs of disks is proportional to the normal force, on scale $aP$
(left) and $F_0$ (right). Red, green, and blue lines distinguish
compressive, tensile, and distant interactions in the cohesive
case, while rattlers appear in grey in the cohesionless sample.
\label{fig:confmethod1}}
\end{figure*}
This state is characterized by its solid fraction
($\Phi=~A^{-1}\sum_i \pi d_i^2/4$) and its coordination number $z$,
defined as the average number of interactions (contacts and
distant attractions) for a particle in the packing, when the
applied pressure is significantly smaller than $F_0/a$ ($P^*\ll
1$), as in the case of small powder samples assembled under
gravity. With the values indicated above (at the end of
Section~\ref{subsec:parameters}) for toner particles,
$F_0/a^2$ (the relevant pressure
scale in 3D) is of the order of $100$~Pa, which corresponds to a
normal consolidation stress in a cohesive powder with $34\%$
solid fraction \cite{VQC04}.

In the absence of
cohesion, the value of the applied pressure does not affect the
properties of the packing (apart from setting the scale of
intergranular forces) provided the typical contact deflection,
$aP/K_N$, is small enough (rigid particle limit). We set this
ratio to the value of $F_0/(aK_N)$ in the cohesive case, \emph{i.e.},
equal to $\kappa^{-1}$ (see table~\ref{tab:param}),
so that typical contact forces are of the same
order of magnitude (due either to $P$ or predominantly to $F_0$)
in both cases.

Effects of the initial solid fraction in the $I$-state, and of
cohesion, friction and rolling resistance parameters on $\Phi$ and $z$
were measured in 3 sets of
samples, with $\Phi_\as=0.13$, 0.36 and $0.52$. Each set consisted of
configurations with the initial disorder (particle radii and
initial positions) abiding by the same probability distribution,
and the same number of particles ($N=1400$). The values of the
friction coefficient, $\mu$, used in these tests were 0.15 and
0.5. The values of $\Phi$ and $z$ in these samples are listed in
tables~\ref{tab:method1-nocoh} and \ref{tab:method1-coh}. Each one
is an average on the different samples, and the indicated
uncertainly is equal to the standard deviation.
\begin{table*}[!htb]
\centering
\begin{tabular}{|c||c|c|c|c|} \cline{2-5}
\multicolumn{1}{c}{} & \multicolumn{4}{|c|}{Non-Cohesive samples} \\ \cline{2-5}
\multicolumn{1}{c}{} & \multicolumn{2}{|c|}{no RR} & \multicolumn{2}{|c|}{RR} \\ \hline
$\Phi_\as$           & $\mu=0.15$ & $\mu=0.5$      & $\mu=0.15$ & $\mu=0.5$    \\ \hline
\multicolumn{5}{c}{Solid fraction}          \\ \hline
$0.130  \pm 0.001$   &
$0.8262\pm 0.0007$   & $0.811\pm 0.001$     &
$0.8238\pm 0.0014$   & $0.803\pm 0.002$     \\ \hline
$0.3631 \pm 0.0006$  &
$0.8256\pm 0.0005$   & $0.811\pm 0.001$     &
$0.8231\pm 0.0013$   & $0.805\pm 0.002$     \\  \hline
$0.5244 \pm 0.0012$  &
$0.8236\pm 0.0007$   & $0.8092\pm 0.0005$   &
$0.8215\pm 0.0005$   & $0.803\pm 0.011$     \\ \hline
\multicolumn{5}{c}{Coordination number}     \\ \hline
$0.130\pm 0.001$     &
$3.174\pm 0.012$     & $2.607\pm 0.022$     &
$3.160\pm 0.024$     & $2.526\pm 0.021$     \\ \hline
$0.3631\pm 0.0006$   &
$3.187\pm 0.025$     & $2.65 \pm 0.02$      &
$3.123\pm 0.013$     & $2.475\pm 0.025$     \\ \hline
$0.5244 \pm 0.0012$  &
$3.181\pm 0.015$     & $2.63 \pm 0.02$      &
$3.15\pm 0.03$       & $2.52\pm 0.02$       \\ \hline
\end{tabular}
\caption{\label{tab:method1-nocoh} Solid fractions and coordination
numbers obtained at the preparation of the specimens in
equilibrated samples under $P/K_N = 10^{-5}$ for non-cohesive
particles, using method 1.}
\end{table*}
\begin{table*}[!htb]
\centering
\begin{tabular}{|c||c|c|c|c|} \cline{2-5}
\multicolumn{1}{c}{} & \multicolumn{4}{|c|}{Cohesive samples} \\ \cline{2-5}
\multicolumn{1}{c}{} & \multicolumn{2}{|c|}{no RR} & \multicolumn{2}{|c|}{RR} \\ \hline
$\Phi_\as$           & $\mu=0.15$ & $\mu=0.5$      & $\mu=0.15$ & $\mu=0.5$    \\ \hline
\multicolumn{5}{c}{Solid fraction}          \\ \hline
$0.130  \pm 0.001$   &
$0.7635 \pm 0.0023$  & $0.751\pm 0.001$     &
$0.757 \pm 0.002$    & $0.709\pm 0.001$     \\ \hline
$0.3631 \pm 0.0006$  &
$0.727\pm 0.001$     & $0.7232\pm 0.0012$   &
$0.710 \pm 0.002$    & $0.688\pm 0.001$     \\ \hline
$0.5244 \pm 0.0012$  &
$0.737\pm 0.002$     & $0.733\pm 0.002$     &
$0.7248\pm 0.0002$   & $0.733\pm 0.002$     \\ \hline
\multicolumn{5}{c}{Coordination number}     \\ \hline
$0.130  \pm 0.001$   &
$3.563\pm 0.005$     & $3.163\pm 0.004$     &
$3.189 \pm 0.008$    & $3.059\pm 0.003$     \\ \hline
$0.3631\pm 0.0006$   &
$3.345\pm 0.009$     & $3.103\pm 0.006$     &
$3.253 \pm 0.003$    & $2.971\pm 0.006$     \\ \hline
$0.5244 \pm 0.0012$  &
$3.189\pm 0.008$     & $3.059\pm 0.003$     &
$3.096 \pm 0.002$    & $2.851\pm 0.001$     \\ \hline
\end{tabular}
\caption{\label{tab:method1-coh} Solid fractions and coordination
numbers obtained at the preparation of the specimens in
equilibrated samples under $P^*=0.01$, $F_0/(K_Na) =  10^{-5}$ for cohesive
particles, using method 1.}
\end{table*}

Tables~\ref{tab:method1-nocoh} and \ref{tab:method1-coh}  show
that the introduction of cohesion reduces the solid fraction at
equilibrium, but this is a limited effect (less than 10\% density
reduction), which is quite insufficient to account for
experimental observations. Unlike powders or clays, 2D particle
packings with $\Phi \ge 0.7$ cannot undergo very large plastic
density increases.

Theses tables also show that the increase of the friction
coefficient and/or the inclusion of rolling resistance in the
model tend to hinder motions and stabilize looser, less
coordinated configurations, which results in a decrease of $\Phi$
and $z$.

However, the observed differences are rather small, especially in
cohesionless systems. To evaluate the influence of RR with a given
value of $\mu$, we define $\Delta X^{(\mu)} = \langle 1 -
X_{\RR}^{(\mu)} / X_{\nRR}^{(\mu)} \rangle$, as the relative
average decrease of the quantity $X$ due to the existence of RR.
For example, for $\mu=0.15$ results differ by a mere
$\Delta\Phi^{(0.15)}=0.24\%$ and $\Delta z^{(0.15)}=1.3\%$, and
for $\mu=0.5$, these variations are $\Delta\Phi^{(0.5)}=0.84\%$
and $\Delta z^{(0.5)}=4.6\%$. Comparing the effect of RR on $\Phi$
with $\mu=0.5$ for $\mu=0.15$ and $\mu=0.5$, one has $
\Delta\Phi^{(0.5)}/\Delta\Phi^{(0.15)} = 3.5$. Likewise, for
coordination numbers $z$, one observes $\Delta z^{(0.5)}/\Delta
z^{(0.15)} \simeq 3.53$. This shows a clear correlation of
variations introduced by friction and RR. The data in the
non-cohesive case also exhibit very little dependence on initial
density $\Phi_\as$.

Results on cohesive systems show similar variations with the
parameters of the contact model (friction and RR), but depend
somewhat more sensitively on $\Phi_\as$.

More refined information on the contact network is provided by the
distribution of local coordination numbers, \emph{i.e.} the
proportions $x_k$ of particles interacting with $k$ neighbors,
for $0\le l\le 6$ (higher values were not observed).
\begin{figure*} [!htb]
   \begin{tabular}{cc}
       \psfig{file=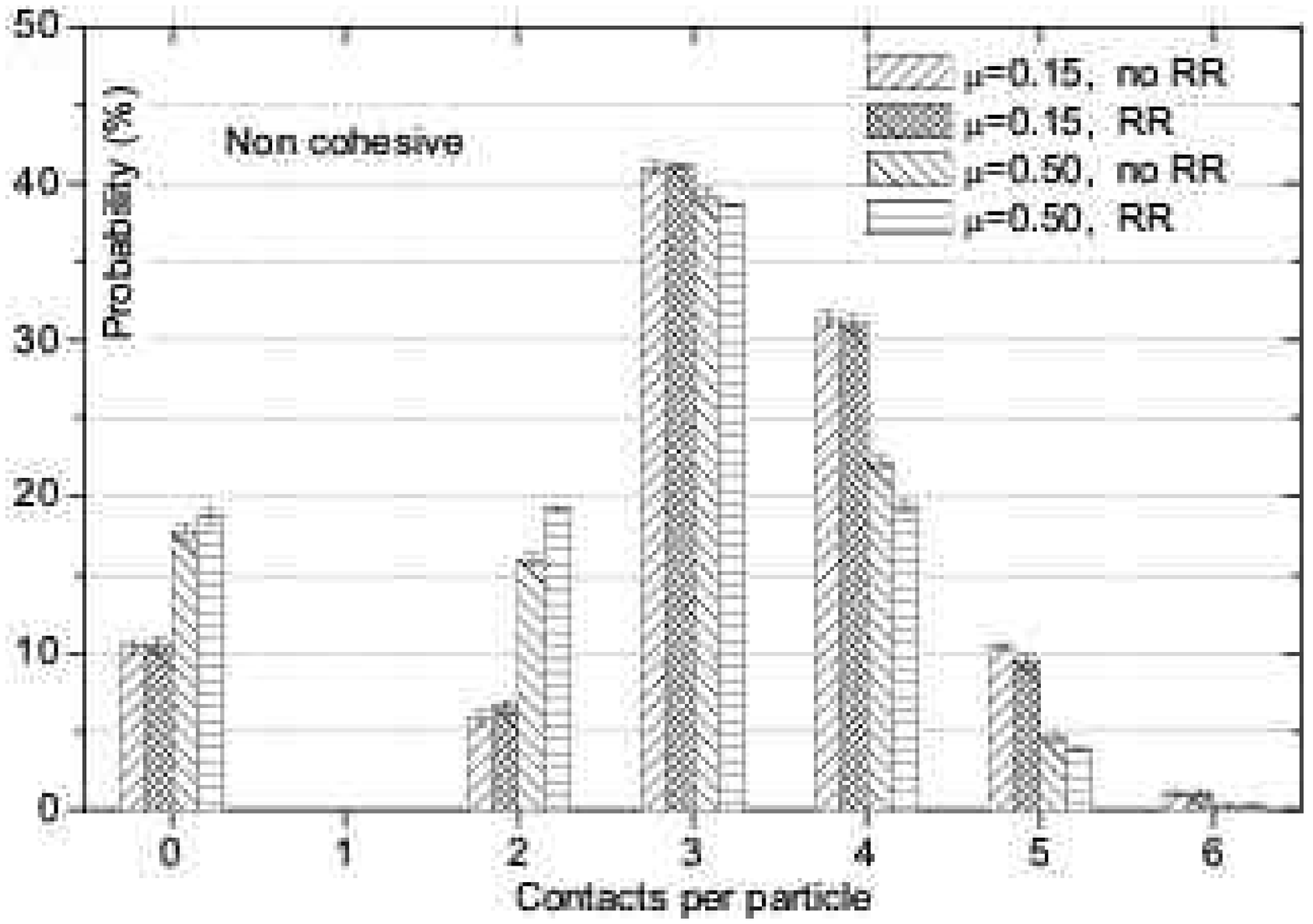,height=5.5cm,angle=0} &
       \psfig{file=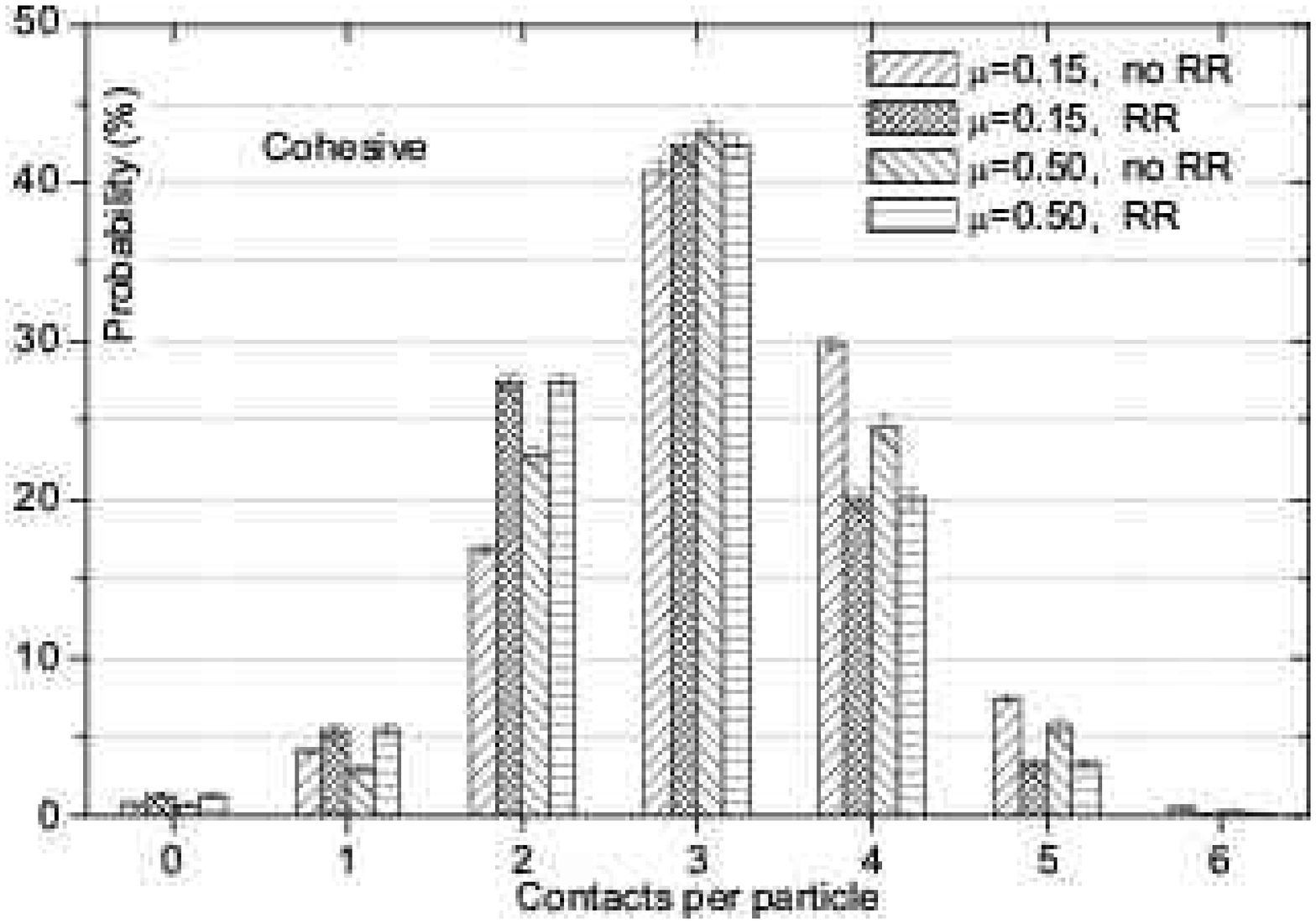,height=5.5cm,angle=0}
   \end{tabular}
   \caption{Distribution of local coordination numbers (percentage
            of total particle number), without (left graph) and
            with (right graph) cohesion.}
   \label{fig:distrib_cont}
\end{figure*}
This distribution is depicted on Figure~\ref{fig:distrib_cont},
for both cohesive and non-cohesive samples. These results gather
information from all the statistically equivalent simulated
samples, and slight corrections were applied in order to ignore
the contacts with ``rattler" particles in the non-cohesive case.
Such particles are those that are free to move within the cage of
their near neighbors, and transmit no force once the system is
equilibrated. If they happen to be in contact with
the \emph{backbone} (\emph{i.e.}, the force-carrying structure),
then the forces carried by such contacts should be below the
tolerance set on the equilibrium requirement, and can safely be
ignored. This is how the population of rattlers is identified. We
observe that it can involve up to 18\% of the total number of
grains in the absence of cohesion (see Fig~\ref{fig:nadm1}).

This contrasts with the cohesive case, for which nearly all the
grains are captured by the force-carrying structure because of
attractive forces, and the rattlers are virtually absent. The
particles with one contact equilibrate when the deflection of that
contact is $h_0$, as defined in~\eqref{eqn:h0}. With RR, such a
particle is entirely fixed. Without RR, it is only free to roll
without sliding on the perimeter of its interacting partner,
because such a contact is able to transmit a tangential force
smaller than or equal to $ \mu  K_Nh_0=\mu F_0 $.

Without cohesion, the coordination of the force-carrying structure
can be characterized with a coordination number $z^*$, different
from $z$: \be z^* = \frac{z}{1-x_0},\label{eqn:defzs}\ee $ z^*$ is
the average number of contacts bearing non-negligible forces per
particle on the backbone. Without cohesion, the backbone (or set
of non-rattler grains) is the rigid part of the packing. With
cohesion and RR, the whole interacting contact network is to be
considered in order to study the rigidity properties of the
system, and there are nearly no particles to eliminate. With
cohesion and no RR, we observe in the samples obtained by the
presently employed procedure (method 1) that the network of
interparticle contacts or interactions is also rigid, apart from
the free rolling of isolated grains with only one contact. (The
rigidity properties of equilibrated samples are discussed below in
Section~\ref{sec:mech} and Appendix~\ref{sec:apprig}).

Cohesive samples in equilibrium also comprise a small number of
pairs of particles interacting without contact, \emph{i.e.}
separated by a gap smaller than the range of attraction, $D_0$.
These are only a small fraction, below 1\%, of interacting pairs.
Such pairs do not contribute to dissipation, since the frictional
and viscous force components are only present in true contacts
between neighboring grains. We observed that the time necessary to
equilibrate the sample tend to increase when such distant
interacting pairs are more numerous.

In addition to the elimination of free rattlers, the
most notable effect of cohesion on local coordination numbers
(Fig.~\ref{fig:distrib_cont}) is to increase the proportion of
disks with 2 contacts. Without cohesion, the Coulomb condition
restricts the angle between the directions of the 2 contacts to
values between $\pi-2\varphi$ and $\pi$, where $\varphi$ is the
friction angle ($\tan \varphi = \mu$). Thus if $\mu$ is small, a
disk with two contacts should have its center close to the line of
centers of its two partners. The increase of the population of
2-coordinated disks as $\mu$ is raised from $0.15$ to $0.5$ (see
Fig.~\ref{fig:distrib_cont}) in cohesionless systems corresponds
to a less severe geometric restriction on contact angles. With
cohesion, contacts may transmit a tangential force reaching $\mu
F_0$ while the normal force is equal to zero. Consequently, a disk
might be in equilibrium with two contact points in arbitrary
positions on its perimeter. As there is no geometric constraint on
the angle between the two contact directions, 2-coordinated disks
are easier to stabilize, and their proportion raises from about
$5\%$ without cohesion to above $15\%$ with cohesion in the case
$\mu=0.15$. A population of disks with one contact (therefore
carrying a vanishing normal force, with deflection $-h=h_0$) is
also present. Those particles are fixed by a small rolling
resistance, but are free to roll on their interacting neighbor
without RR. Such a rolling motion is not damped in our model.
Therefore, on waiting long enough, they should eventually stop
after a collision, in a stable position with 2 contacts. Such a
collision is bound to happen because the contact network is
completely connected. However, we stop our calculations when the
kinetic energy is below a set tolerance (see
Section~\ref{subsec:equilibrium}), and we do not wait until all
freely rolling disks reach their final position. Hence the
remaining population of disks with one contact in samples without
RR.

The final configuration, with this preparation method, depends
somewhat on the rate of compaction in the assembling stage. The
latter is related to the choice of the dynamical parameter $M$,
the ``mass" with which the changes in cell dimensions are computed
with Eqn.~\eqref{eqn:cell}. The slight influence of the  initial
solid fraction, $\Phi_\as$, also relates to such dynamical
effects: a lower value of $\Phi_\as$ entails larger colliding
velocities, which favors larger final solid fractions.

Although some of the aspects of the model (in particular the
homogeneous shrinking imposed through the periodic cell dimensions
in a dynamical regime) do not correspond to experimental
conditions, configurations of type 1 should be regarded as typical
results of fast assembling processes, in which the particles are
requested to balance the external pressure before stable loose
structures can be built. When the toner particles mentioned
at the end of Section~\ref{subsec:parameters} are first fluidized,
and then settle under their own weight, a rough estimate of the
settling time, assuming particles are settling individually in
air, and fall over distances of order 1~cm, is $\sim 1$~s.
Fig.~\ref{fig:prep}, with the value $T_0 \sim 10^{-5}$~s corresponding to
such particles, shows that
the duration of the ``method 1'' compression process is a few milliseconds.
In practice, due to the presence of the surrounding fluid, the
packing of a powder in a loose state
by settling and compaction of an initially fluidized
state is therefore considerably slower than this numerical process.

In the next section, we consequently turn to the opposite limit,
in which the external confining pressure is felt only after large,
tenuous contact networks are formed.

\subsection{Method 2 \label{sec:method2}}
\subsubsection{Numerical procedure.}
\begin{figure*}[!htb]
\resizebox{0.98\textwidth}{!}{\includegraphics*{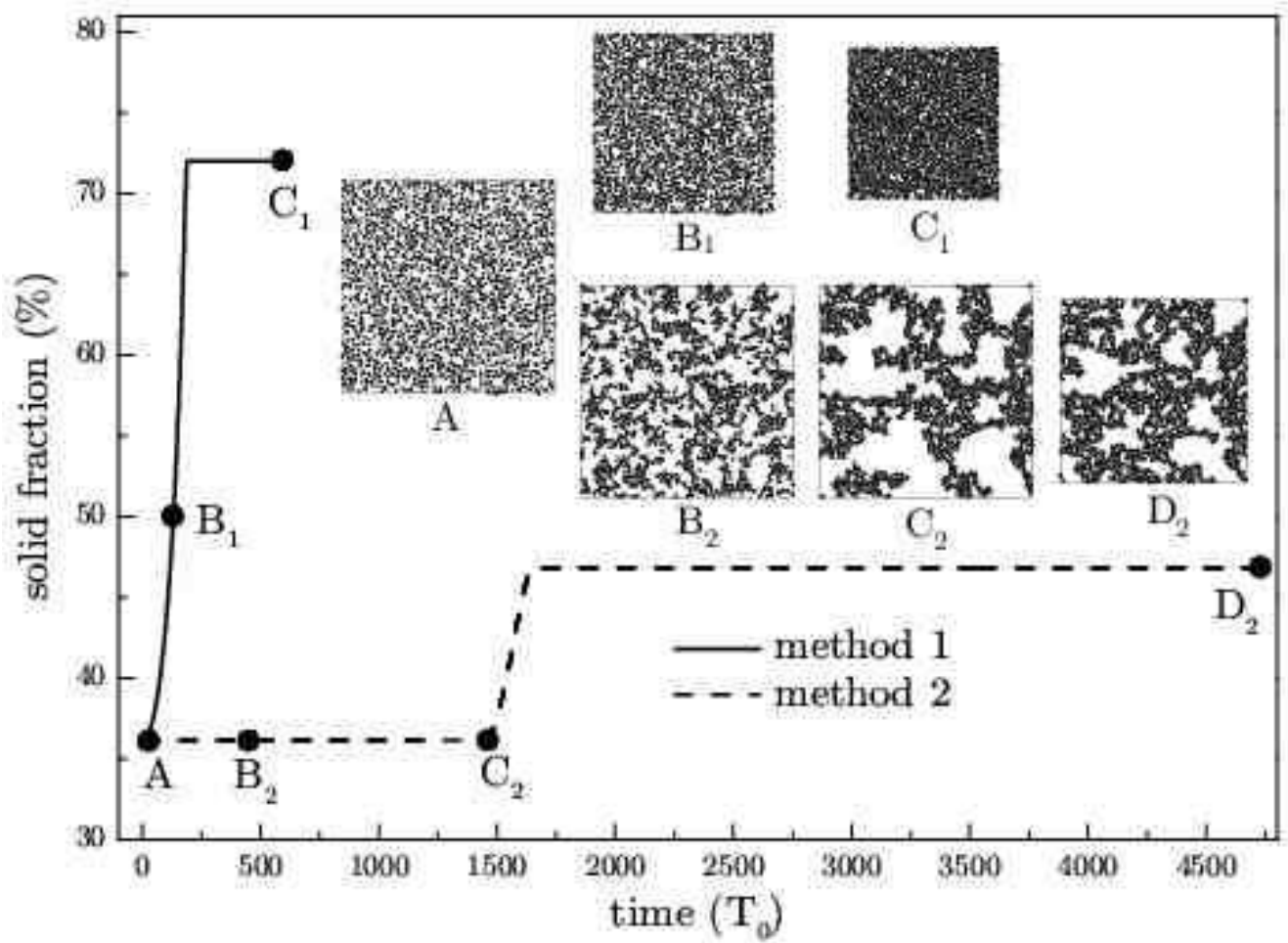}}
     \caption{\label{fig:prep}
Solid fraction versus time for both preparation procedures,
showing some aspects of the configurations at different stages.
Point A is the initial state (or $\Phi_\as$). Aspects of
configurations are shown for intermediate states B$_1$ and B$_2$,
and for final equilibrated states C$_1$ and D$_2$ (at $P^*=0.01$).
Point C$_2$ corresponds to the stage when all disks are assembled
in a unique aggregate, then equilibrated at $P^*=0$ (both
aggregation and equilibration stages take place between A and
C$_2$). The time unit is $T_0={\disty \sqrt{ma/F_0}}$. Note the
duration of the preparation process with method 2, and the
difference in final equilibrated states compared to method 1.}
\end{figure*}
The second method to prepare numerical samples allows for
aggregate formation before imposing an external pressure. Along
with method 1, its different stages are schematically presented on
Fig.~\ref{fig:prep}.

The aggregation phenomenon plays an important role in the
experimental preparation procedure of refs.~\cite{VCQ01,VCQ201} in
which powder particles in a fluidized bed collide and stick to
each other. Then they settle under their weight when the upwards
air flow is abruptly shut off. The numerical method was designed
to reproduce, in some idealized way, the final state of a set of
colliding particles in the absence of external force fields. In
the initial disordered low-density configuration (the same
$I$-state as in method 1), particles are now attributed random
velocities drawn according to a Maxwell distribution, with mean
quadratic velocity $V_0$.

We performed systematic sets of simulations of disk packings with
$V_0 = 9.48 V^*$ (see Eqn.~\eqref{eqn:vsep}). $V_0$ is thus large
enough for the initial kinetic energy to overcome potential energy
barriers in the process of aggregation. (The dependence of the
final packing structure on this initial velocity of
\emph{agitation}, or ``granular temperature", in the assembling
stage in systems with small RR will be studied in
Section~\ref{sec:vinit}).

Once launched with such random velocities the particles are left
to interact and stick to one another within a cell of constant
size, forming larger and larger aggregates, as appears on the
image marked ``B2" on Fig.~\ref{fig:prep}. Eventually, all
particles are connected to one another by adhesive contacts, and
reach an equilibrium position. At this stage, the two degrees of
freedom of the cell are set free, and the stress-controlled
calculation proceeds with $\sigma_{11}=\sigma_{22}=0$ (or $P^*=0$)
until an equilibrium state is reached. This relaxation step does
not lead to any rearrangement of the contact structure, it only
entails a very small increase of the solid fraction (hence the
values slightly larger than $\Phi_\as$ given below). The final
equilibrium structure exhibits large density inhomogeneities, as
apparent on Fig.~\ref{fig:prep}, which are characteristic of
aggregation processes~\cite{SMI90}, and will be quantitatively
studied in Section~\ref{sec:geometry}.

Unlike cohesionless systems, which are devoid of any ``natural"
state of stress, clusters of cohesive particles can exist in a
well defined state of mechanical equilibrium in the absence of any
external force. Once the state at zero pressure is obtained, we
subsequently apply the same load $P^*=0.01$ as in method 1, which
results in further compression and notable changes in the packing
structure: $\Phi$ increases from values close to $\Phi_{\as}$ up
to the $0.45$--$0.55$ range (see fig.~\ref{fig:prep}).
Nevertheless, the final solid fraction under $P^*=0.01$ is
considerably lower than the one obtained with method 1.

It should be noted on Fig.~\ref{fig:prep}, which summarizes the
assembling procedures, that the aggregation stage makes method 2
computationally quite costly because of the time necessary for
clusters to merge, and especially for the stabilization of loose
samples in equilibrium configurations (lower contact numbers
implying lower rates of energy loss as well as larger and slower
fluctuations of soft, tenuous structures).
In an attempt to limit the influence of compaction dynamics, which
results in denser samples when the lower density of the initial
state allows the compaction process to accelerate more (as noted
in Sec.~\ref{sec:method1}), we tested the effect of limiting the
maximum strain rate $\dot{\epsilon}_{\text{max}}$. Without any
limitation, we obtained a maximum value $\dot{\epsilon} \simeq
0.15~T_0^{-1}$. Using the samples with $\Phi_\as = 0.13$ (the
lowest value used in this work) with $K_N=10^5 F_0/a$, three
different values for $\dot{\epsilon}_{\text{max}}$ were tested:
$0.10~T_0^{-1}$, $0.05~T_0^{-1}$ and $0.015~T_0^{-1}$. The
condition $\dot{\epsilon}\ge 0.10~T_0^{-1}$ gave a final state
close to the original one. The others two values produced similar
results, with a relative decrease in density of about $10\%$
compared to the original procedure. We chose to enforce condition
$\dot{\epsilon}_{\text{max}} = 0.05~T_0^{-1}$, to save
computational time. This value has been applied to prepare all
samples studied in the following.

Fig.~\ref{fig:prep} shows that
method 2 succeeds in stabilizing open structures. Final solid fractions
agree with the typical values observed in powders
if one uses the correspondence between 2D and 3D packing fractions
suggested by Campbell in~\cite{Campbell85}:
\be
\Phi_{\mbox{\tiny 3D}} = \frac{4}{3\sqrt{\pi}}
\Phi_{\mbox{\tiny 2D}}^{3/2} \simeq 0.752 \Phi_{\mbox{\tiny
2D}}^{3/2}.
\ee
Numerical samples under $P^*=0.01$, with solid fractions around $45\%$, would correspond to a
powder consolidated in the laboratory under 1~Pa with a solid
fraction of about $23\%$. This is in satisfactory agreement with
the experimental results of Ref.~\cite{VQC04}.

We therefore regard method 2 as an appropriate way to reach an
essential objective of this work, since stable loose structures
are obtained.

Although we perform simulations of
a \emph{mechanical} model, the final configurations exhibit at first
sight (Fig.~\ref{fig:prep}) similar features as those obtained
with \emph{geometric} algorithms implemented in numerical studies
of colloid aggregation models~\cite{MEA99,KBBW03}.
We are not aware of similar results in the literature, at least
with equilibrium requirements comparable to those of
Sec.~\ref{subsec:equilibrium}.

Tenuous, fractal-like contact networks contain
denser regions and large cavities. Such heterogeneities produce
long range density correlations, to be analyzed in
Sec.~\ref{sec:geometry}. Without tensile contact forces, the walls
of the cavities, comprising particles that are pushed towards the
hole by the resultant of contact normal forces, would tend to
buckle in.

{\em We regard method 2 as yielding typical results for assembling
processes in which particles form tenuous aggregates before they
are packed in a structure that is able to sustain a confining
stress}. In the sequel, we focus on the tenuous structures
obtained with method 2.

\subsubsection{Global characterization of loose packings at $P^*=0$ and $P^*=0.01$.}
We simulated four samples with 1400 disks and three of 5600 for
$\Phi_\as = 0.36$, rather than lower initial densities, in order to achieve statistical significance at
affordable computational costs, and to check for possible size effects. This set of samples will be
denoted as series A.

Samples with $\Phi=0.13$ (series A0), which require
the initial cell to shrink more before a stable network can resist the pressure, request
longer calculations. Although some samples were prepared at $P^*=0$, we do not use them any more in the
following, except for the values showed in Table~\ref{tab:p0}.

To accelerate the numerical assembling procedure,
we also created samples with $K_N =10^4 F_0/a $,
using an intermediate value of $\Phi_\as=0.26$, and softer contacts, such that $\kappa=10^2$
in the initial aggregation stage (recall the time step is proportional to $A/\sqrt{K_N}$).
Once equilibrium was reached with $P^*=0$, we slowly changed the stiffness parameter
from $\kappa=10^2$ to $\kappa=10^4$, and recorded the final equilibrated
configuration. This procedure is about ten times
as fast as the normal one, and generates similar structures
and coordination numbers as series A prepared with the same
$I$-state density. We shall refer to this set as series B.

In table \ref{tab:p0} we
list the corresponding results for solid fractions and coordination
numbers. In such data we did not find a significant difference
between the two different sample sizes, and therefore we did not
distinguish between sizes in the presentation of statistical
results.
\begin{table*}[htb]
\centering
\begin{tabular}{|c|c|c|c|c|} \cline{2-5}
\multicolumn{1}{c}{}         & \multicolumn{2}{|c|}{no RR}           & \multicolumn{2}{|c|}{RR}            \\ \hline
$\Phi_\as$ ($z=0$)           & $\Phi$            & $z$               & $\Phi$            & $z$             \\ \hline
$0.1301\pm0.0003$ (series A0) & $0.1303\pm0.0003$ & $3.197\pm0.002$   & $0.1304\pm0.0003$ & $2.656\pm0.007$ \\ \hline
$0.2649\pm0.0006$ (series B)  & $0.265\pm 0.001$  & $3.123\pm 0.004$  & $0.2652\pm0.023$  & $2.963\pm0.006$ \\ \hline
$0.361\pm0.007$ (series A)   & $0.3616\pm0.0003$ & $3.1407\pm0.0016$ & $0.361\pm0.009$   & $2.660\pm0.004$ \\ \hline
\end{tabular}
\caption{\label{tab:p0} Values of $\Phi$ and $z$ on equilibrating
configurations at $P^*=0$ with $\mu=0.5$. Samples with $\Phi_\as=0.26$
correspond to $\kappa=10^4$, the others to $\kappa=10^5$.}
\end{table*}
The tenuous networks obtained with method 2 collapse on changing
the pressure: Table~\ref{tab:p001} gives the new values of $\Phi$
and $z$ after the compaction caused by the pressure increase from
$P^*=0$ to $P^*=0.01$.

\begin{table*}[htb]
\centering
\begin{tabular}{|c|c|c|c|c|} \cline{2-5}
\multicolumn{1}{c}{}       & \multicolumn{2}{|c|}{no RR}           & \multicolumn{2}{|c|}{RR}            \\ \hline
$\Phi_\as$ ($z=0$)         & $\Phi$            & $z$               & $\Phi$            & $z$             \\ \hline
$0.2649\pm0.0006$ (series B)     & $0.448\pm0.006$   & $3.235\pm0.003$   & $0.42\pm0.01$     & $3.085\pm0.005$ \\ \hline
$0.361\pm0.007$ (series A) & $0.472\pm0.008$   & $3.175\pm0.003$   & $0.524\pm0.008$   & $2.973\pm0.004$ \\ \hline
\end{tabular}
\caption{\label{tab:p001} Values of $\Phi$ and $z$ in equilibrated
configurations at $P^*=0.01$. These results are averaged over the
whole set of samples prepared with $\mu = 0.5$, for
$\Phi_\as=0.26$ (series B) on the one hand, and for
$\Phi_\as=0.36$ (series A) on the other hand. Series A0, prepared
with $\Phi_\as=0.13$, yielded very similar results but due to
computational costs the number of samples was too small to record
data in statistical form. }
\end{table*}
Structural changes between $P^*=0$ and $P^*=0.01$ are shown on
Fig.~\ref{fig:p0}, which illustrates by means of four selected
snapshots the mechanism of the closing of pores in a 1400 disks
sample of series A. The first image corresponds to equilibrium at
$P^*=0$, and the fourth one to equilibrium under $P^*=0.01$. The
two others show intermediate, out of equilibrium configurations
during the collapse. One may appreciate how the denser regions
grow and merge while pores shrink.
\begin{figure*}[!htb]
  \begin{center}
     \psfig{file=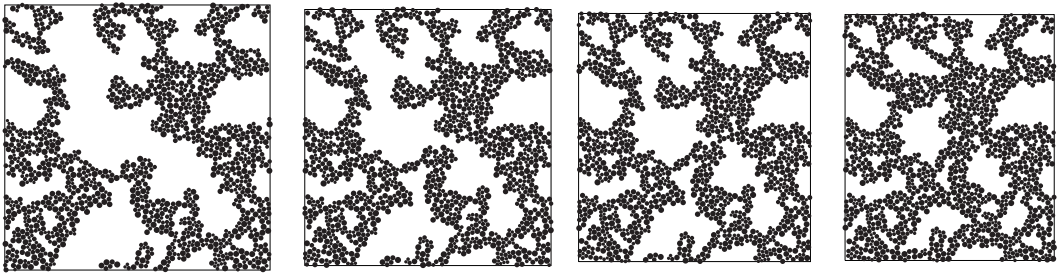,height=4.cm,angle=0}
     \caption{Configuration of 1400 disk sample of series A without rolling resistance.
              Note the gradual closing of pores as the external pressure is increased
              from $P^*=0$ (first image) to $P^*=0.01$ (last image) going through two intermediate stages.
     \label{fig:p0}}
   \end{center}
\end{figure*}
Fig.~\ref{fig:p0} also makes it quite evident that the size of
1400 disk samples is not very much larger than the scale $\xi$ of
density heterogeneities (typical diameter of large pores or dense
regions, which will be studied in Sec.~\ref{sec:geometry}). These
systems will exhibit large fluctuations in their mechanical
properties: the rectangular shape of the final configuration
displayed on Fig.~\ref{fig:p0} shows that the disorder is large
enough for the mechanical response of the system to become
anisotropic. Isotropy should be recovered in the limit of large
sample sizes, $L\gg \xi$.

Finally, Fig.~\ref{fig:distrib_cont_m2} displays the histogram of
local coordination numbers (percentage of particles interacting
with $k$ others, $0\le k\le 6$), for the same samples as those of
Tables~\ref{tab:p0} and \ref{tab:p001} ($\mu=0.5$,
$\Phi_{\as}=0.36$).
\begin{figure} [!htb]
       \psfig{file=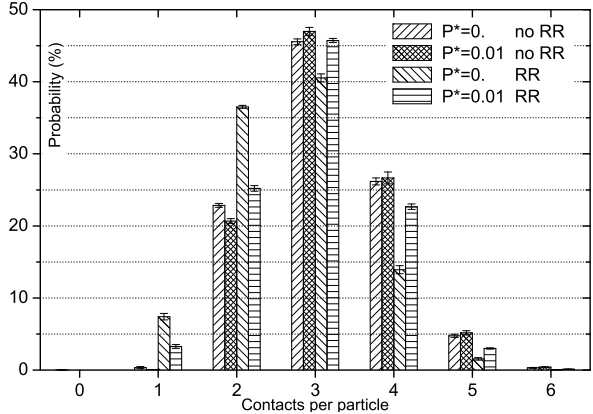,height=5.5cm}
   \caption{Distribution of local coordination numbers in loose samples
   of series A obtained with method 2.
   Samples of series B gave a similar distribution.
   \label{fig:distrib_cont_m2}}
\end{figure}
It is remarkable that this distribution, in spite of the large
difference in sample geometries, remains rather close to the one
observed in the denser packings made with method 1 (compare
$P^*=0.01$, $\mu=0.5$ case), just like global coordination numbers
take very similar values in samples prepared with both methods
(see Tables~\ref{tab:method1-coh} and \ref{tab:p001}), in spite of
the very different solid fractions.

An essential conclusion of the present study is therefore, for one
given material, \emph{the absence of a general relation between
the density of a cohesive packing and its coordination number}, in
spite of previous  claims~\cite{YZY03a}. Both quantities are
determined, rather, by the conjunction of micromechanical laws and
sample preparation history.

\subsubsection{Effects of micromechanical parameters}
Adhesion should enhance the role of sliding friction and rolling
friction, because the limiting values for tangential contact
forces and rolling moments are both proportional to the elastic
repulsive part of the normal force, $N^e$ ( $|T_{ij}| < \mu
N_{ij}^e$, $|\vec{\Gamma}^{~r}_{ij}| < \mu_r N_{ij}^e$).
Consequently, contacts with the equilibrium value $h_0$ of the
elastic deflection for an isolated pair of grains transmit no
normal force, but are able to sustain tangential force components
as large as $\mu F_0$ and rolling moments as large as $\mu_r F_0$.
Those values might turn out to be large in comparison to the
typical level of intergranular forces under low external pressure
($P^* \ll 1$).
\begin{figure*}[!ht]
   \begin{tabular}{cc}
       \psfig{file=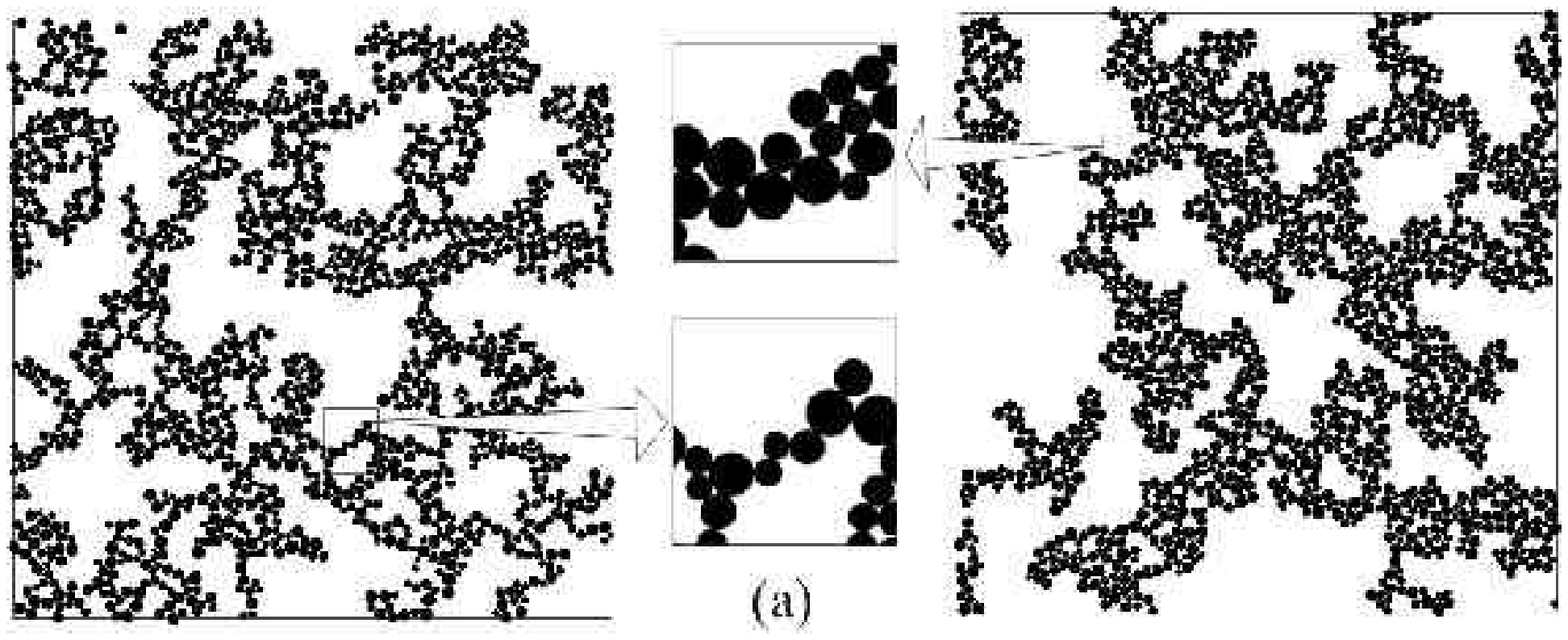,height=4.5cm,angle=0} \\
       \psfig{file=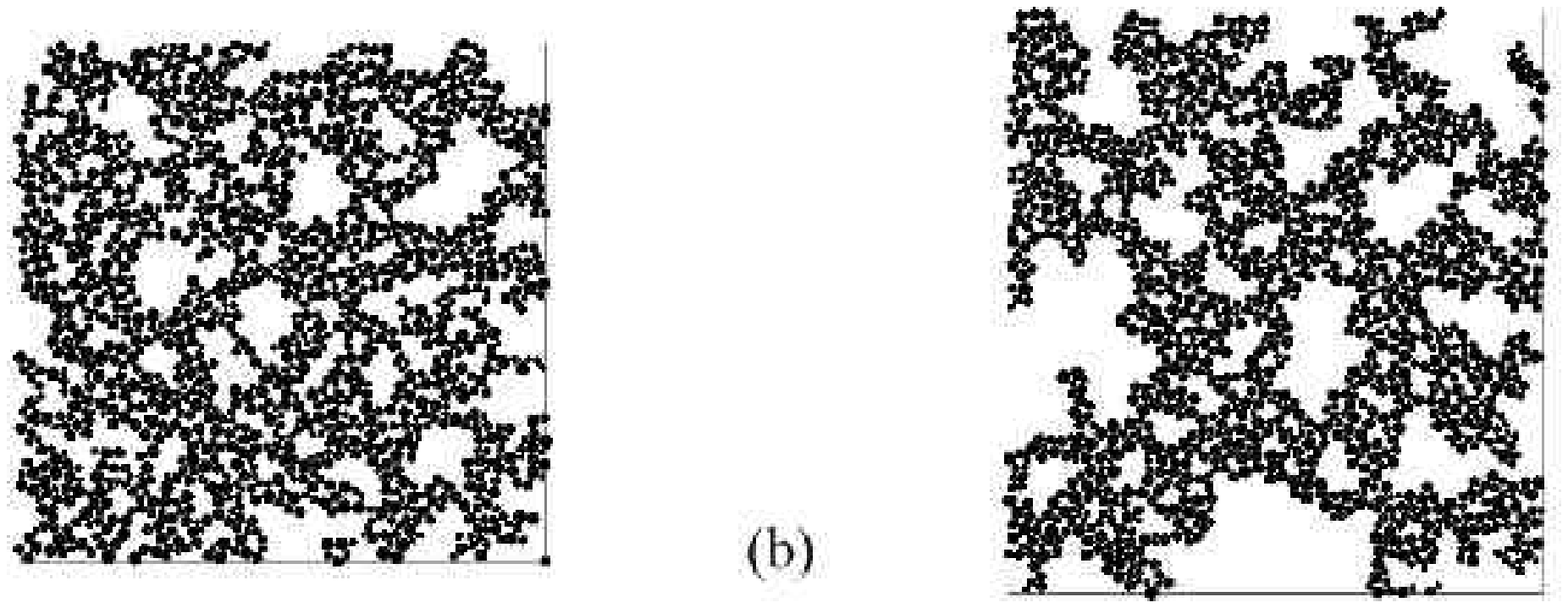,height=4.5cm,angle=0}
   \end{tabular}
   \caption{Typical configurations of 1400 disk
samples of series A with (left) and without (right) rolling
resistance, at $P^*=0$ (a) and $P^*=0.01$ (b). Note the difference
in local structure of thin ``beams'' joining dense regions with or
without RR. \label{fig:chains}}
\end{figure*}
Therefore, even very low values of $\mu$ and $\mu_r$ should affect
the final structure of equilibrated packings considerably more
than in the cohesionless case. This is indeed the case for the
coordination numbers observed in our simulations (see tables
\ref{tab:method1-nocoh} and \ref{tab:method1-coh}) which dropped
more significantly, upon introducing the small level of RR we have
been using, in cohesive systems than in cohesionless ones.

On Fig.~\ref{fig:chains} we show the configurations at $P^*=0$ (a) and
$P^*=0.01$ (b) of the \emph{same} sample assembled using method 2
with RR (left) and without RR (right).
The denser regions in the inhomogeneous packings are joined by
slender ``arms" (see Figs.~\ref{fig:p0} and \ref{fig:chains}).
Such arms can in principle reduce to a chain of particles in the
presence of rolling resistance. Such chains are otherwise destabilized by a rolling
mechanism, hence the difference in the thickness of the arms with
or without RR (see the blown-up detail in
Fig.~\ref{fig:chains}-a), the lower coordination numbers of
configurations assembled with RR (Tables~\ref{tab:p0} and
\ref{tab:p001}). This might also explain the greater fragility of
equilibrium configurations with RR, in which a larger compaction
step (see Table~\ref{tab:p001}) is necessary, on applying
$P^*=0.01$, before a new stable structure is reached.

\begin{figure}[!hb]
\centering
\includegraphics[width=1.0\columnwidth]{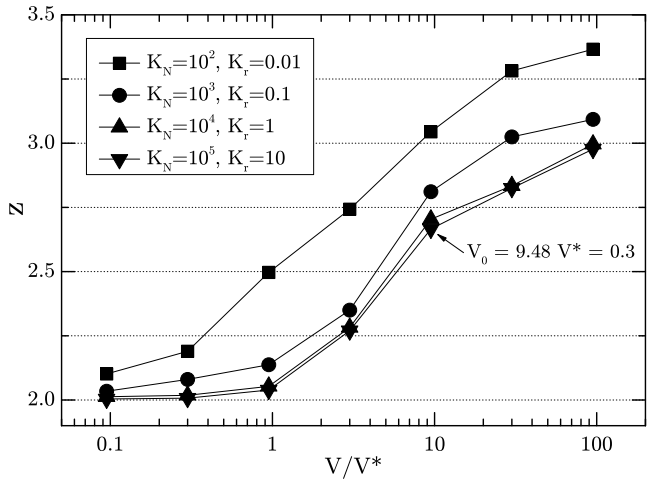}
\caption{\label{fig:z-v0} Final coordination number $z$ versus
initial quadratic average velocity in agitation stage of method 2,
normalized by characteristic velocity $V^*$. The arrow points to
the value most often used in our calculations. }
\end{figure}
Another important parameter is the initial velocity of agitation, $V_0$.
Its influence has been assessed on one 1400 disks sample,
with $\Phi_\as=0.36$. The changes of coordination number with $V_0$
at $P^*=0$ are presented on Fig.~\ref{fig:z-v0}.

Low velocity values produce more tenuous aggregates ($z\sim 2$),
since even a small level of RR is able to slow down local
rearrangements and stabilize tree-like structures (\emph{i.e.,}
devoid of flops) immediately after the collisions between
particles or small clusters.

A large kinetic energy cannot be absorbed by the RR, and as a
result disks are able to rotate, which leads to better connected
structures ($z\sim 3$).  In a sense, a large $V_0$ kills the
effects of RR, and packings are similar to those made without RR
in such cases.

We therefore conclude that the connectivity of loose samples with
RR assembled by aggregation depends on the initial magnitude of velocity
fluctuations and on the level of rolling friction.

As figure \ref{fig:z-v0} shows, the same trend was found on
reducing contact stiffness parameter $\kappa$, as a larger
translational and rotational compliance creates more contacts.

$V_0$ is analogous to the particle fluctuating velocity in
experiments on gas-fluidized beds of xerographic toners under
gravity~\cite{VCQ01}. Such velocities are larger than the gas
velocity by two orders of magnitude. Typically, one has
$v_{\mbox{\tiny gas}} \sim 1 - 4$ mm/s, while $V^*$, deduced from
the contact parameters with relation~\eqref{eqn:vsep} is about
1~cm/s. Such a value is therefore comparable to the particle
fluctuation velocity.

Of course, such a comparison is only indicative, because the
influence of $V_0$ on packing structures depends on $\mu_r$, and
is also very likely to be affected to some extent by the viscous
dissipation model we have adopted. Both rolling resistance and
viscous forces are micromechanical features for which no accurate
physical identification is available. Yet, it seems plausible that
powder packings, because of their initial agitated states,
stabilize in better connected states than predicted by geometric
aggregation models.

We now turn our attention in the next section to the forces in the
contact networks, in particular the loose ones formed with method 2.
\section{Mechanical characterization of contact networks \label{sec:mech}}
Many numerical studies, in the past 15 years, have addressed the
issue of contact network geometry and force distribution in
\emph{cohesionless} systems~\cite{RJMR96}. The image of force
chains, \emph{i.e.} a pattern in which larger intergranular forces
tend to line up on the scale of several grains, was evidenced in
experiments~\cite{MJN98,BMMJN01} and
simulations~\cite{OR97b,MJS00}, and the p.d.f of contact force
values has often been measured and studied. An interpretation of
the mechanical role of ``force chains''~\cite{RWJM98} is that they
carry the essential part of deviatoric stress, while the contacts
carrying the lower forces are less sensitive to stress orientation
and laterally stabilize the strong force chains against buckling.

The main features of the distribution of forces and their spatial
correlations have been reproduced by approximate
models~\cite{CLMNW96} based on local equilibrium rules on each
grain, supplemented by inequality constraints. One important such
constraint is released in cohesive systems, in which normal force
components can have either sign. It is therefore worth
investigating how the usual features of force-carrying structures
in equilibrated granular packings are affected by the presence of
negative normal forces. One may also wonder to what extent the
considerable difference in the density fields will affect the
force patterns, given that the coordination of the force networks,
as observed previously, does not seem to be very sensitive to
density levels and density fluctuations.
\subsection{Force scale and force distribution}
The first obvious distinction between cohesive and non-cohesive
systems is the appearance of a new force scale $F_0$, in addition
to the one provided by the confining pressure, \emph{i.e.}, $aP$,
the ratio of those characteristic forces defining the reduced
pressure, $P^*$. It is especially interesting to investigate the
values and spatial organization of forces in systems with $P^*\ll
1$, as little information is to be found in the literature on this
issue: numerical studies of loose cohesive
systems~\cite{WoUnKaBr05} tend to focus on density and geometry of
packings as a function of applied stresses. Some information on
force networks is provided in a recent publication~\cite{RiEYRa06}
on bead assemblies with capillary cohesion, but the confining
stress is considerably higher is that study ($P^*$ of order 1)
than in the present one.

In the absence of cohesion, the distribution of force values is
usually presented in a form normalized by its average, which
itself scales with the applied pressure. This scaling can be made
more quantitative on using a general relation between pressure $P$
and the average normal contact force $F_N\defeq\ave{N_{ij}}$ and
particle diameter $d$, which is known in the literature on powders
as the Rumpf formula. We write it here in a form involving the
spatial dimension $D$, which is valid both for $D=2$ and $D=3$:
\be  \label{eqn:rumpf1}
P = \frac{1}{\pi} \frac{z \Phi  }{d^{D-1}}F_N.
\ee
In~\eqref{eqn:rumpf1}, $d$ stands for the typical grain diameter.
This relation can be made more accurate if one notes that it stems
from the standard formula for stresses in an equilibrium
configuration (see the r.h.s. term in Eqn.~\ref{eqn:cell}). To
derive the formula, defining $P =
\frac{1}{D}\sum_{\alpha=1}^D\sigma_{\alpha\alpha}$, the average
pressure, one assumes $h_{ij} \ll R_i+R_j$ and then neglects
correlations between particle radii and forces, assuming
\be
    \ave{N_{ij}(R_i+R_j)} \simeq  F_N \ave{d}. \label{eqn:corrneg}
\ee
Then, with a simple transformation of the sum, one obtains
\be  \label{eqn:rumpf_D}
    P = \frac{1}{\pi} \frac{\ave{d}}{\ave{d^D}} z \Phi F_N.
\ee
With $D=2$ and our diameter distribution (for which $\ave{d}=3a/4$ and
$\ave{d^2}=7a^2/12$) this yields
\be
    F_N = \frac{7 \pi a}{9} \frac{P}{z \Phi}.
\label{eqn:rumpf2D}
\ee
We found relation~\eqref{eqn:rumpf2D} to be remarkably accurate in
all our simulations, with or without cohesion, with configurations
obtained by either method 1 or method 2, thereby checking that the
correlations between particle sizes and contact forces could
safely be neglected on writing~\eqref{eqn:corrneg}.

Without cohesion, Eqn.~\eqref{eqn:rumpf2D} yields the correct
scale  for forces, \emph{i.e.} the frequency of occurrence of
intergranular forces larger than a few times $F_N$ is very small.
With cohesion, when $P^*=0$ or $P^*\ll 1$, contact forces of order
$F_0$ are quite common, as shown on Fig.~\ref{fig:distrib_forces},
on which normal force distributions are represented.
\begin{figure}[!htp]
\subfigure[No cohesion, $K_N/P=10^5$]{
\includegraphics*[width=0.9\columnwidth]{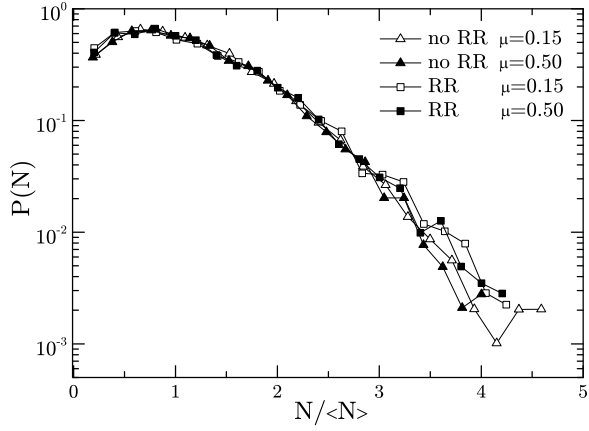}
\label{fig:distfnadm1}}
\subfigure[With cohesion, method 1, $P^*=0.01$, $\mu=0.15$ and
$\mu=0.5$]{
\includegraphics*[width=0.9\columnwidth]{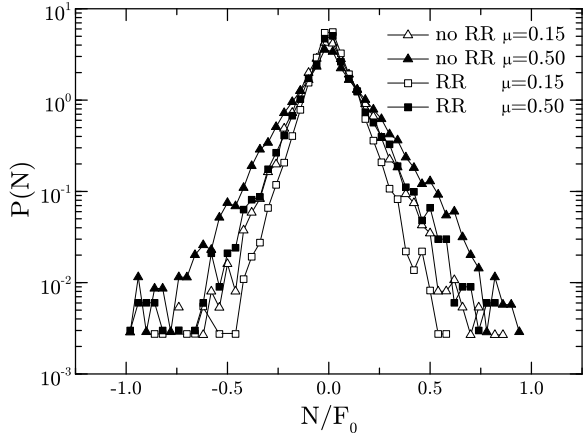}
\label{fig:distfadm1}}
\subfigure[With cohesion, method 2, $P^*=0$ and $P^*=0.01$, $\mu=0.5$]{
\includegraphics*[width=0.9\columnwidth]{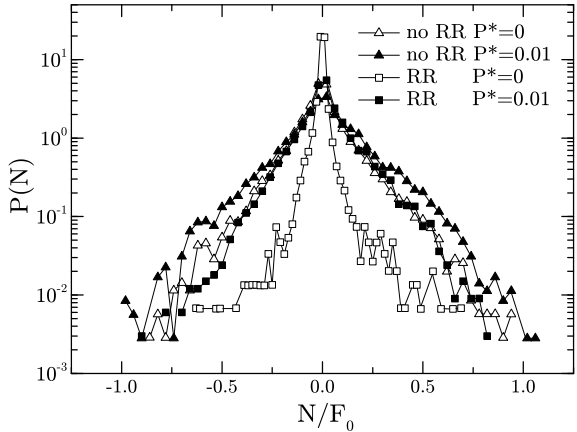}
\label{fig:distfadm2}} \caption{Distribution of normal forces for
series A samples. The non-cohesive case (\ref{fig:distfnadm1}) is
normalized by the average repulsive elastic part. The cohesive
cases (\ref{fig:distfadm1} and \ref{fig:distfadm2}) are normalized
by $F_0$ (note that the average of the elastic part of $N$ is
$\ave{N_e}\simeq F_0$ in cohesive cases with $P^*\ll 1$).
\label{fig:distrib_forces}}
\end{figure}
Hence Eqn.~\eqref{eqn:rumpf2D} cannot be used to predict
``typical" contact forces. The presence of forces of order $F_0$
explains the sensitivity of type 1 and type 2 samples with $P^*\ll
1$ to friction coefficient and rolling resistance: densities and
coordination numbers (tables~\ref{tab:method1-coh}, \ref{tab:p0}
and \ref{tab:p001}), in cohesive systems prepared under $P^*=0$ or
$P^*=0.01$ with $\mu=0.15$ and with $\mu=0.5$, or with and without
RR, differ significantly. Otherwise, if contact forces were of
order of the average $F_N$, the value of which is correctly predicted
by~\eqref{eqn:rumpf2D}, thresholds $\mu F_0$ or even $\mu_r F_0$
would be very large compared to typical forces and moments, and
become irrelevant.

(It should be recalled that Rumpf's name is often associated (as in
Ref.~\cite{RiEYRa06}) to a means to predict the macroscopic
tensile  strength of a powder. As the essential ingredient of the
Rumpf approach~\cite{Rumpf} is Eqn.~\ref{eqn:rumpf1}, we refer
here to that relation (like in~\cite{C05}), as the \emph{Rumpf
formula}).

Normal force distributions in cohesionless, cohesive type 1 and cohesive type 2 samples,
the latter being obtained with $\Phi_\as=0.36$ (series A), are shown on
Fig.~\ref{fig:distrib_forces}. Those distribution functions are
roughly symmetric about $0$, decay approximately exponentially at
intermediate values, and vanish  at $-F_0$, and $F_0$. In type 2
samples without RR, for $P^*=0$, there is a finite proportion of
contacts carrying vanishing forces, about one fourth in the A
series ($\Phi=0.36$). In addition to this Dirac mass, there might
be a power-law divergence near 0, with an exponent our level of
statistics  is not sufficient to resolve accurately (about 0.6 to
0.8 in the range of forces between $10^{-3}F_0$ and $10^{-2}F_0$).
This proportion of zero forces is smaller, down to $9\%$, with RR,
and drops as $P^*$ reaches $0.01$, to $7\%$ and $3 \%$,
respectively, without and with RR. It is worth pointing out that
the corresponding contacts carry zero \emph{total} forces,
\emph{i.e.} vanishing normal components ($-h=h_0$,
see~\eqref{eqn:h0}) and no tangential elastic displacement either.
In principle we cannot distinguish them from forces below the
numerical tolerance defined in Sec.~\ref{subsec:equilibrium}.
However, as we shall argue below in Section~\ref{subsec:forces},
under $P^*=0$ one could expect all contact forces to vanish, and
non-zero forces are related to the small, but finite degree of
force indeterminacy.

Before turning our attention to such features and to the spatial
organization of forces, let us briefly discuss the differences
between sets of (type 2) samples A and B. B samples, which are
obtained with the ``accelerated'' procedure and $\Phi_\as=0.26$
(see Sec.~\ref{sec:preparation}), exhibit, due to their specific
history, larger forces at $P^*=0$, with as many as $10\%$ of
the contacts transmitting normal forces $N$ such that
$\abs{N}>F_0/10$, while this proportion lies below $2\%$ in A
samples. On the other hand, B samples are looser,
with more open contact networks under $P=0$ and a
larger proportion of contacts (about one third in
configurations without RR) carrying vanishing forces. In the
following we shall use them to illustrate qualitative tendencies
in very loose samples.

When the pressure is increased to $P^*=0.01$, differences in force
distributions between A and B samples, despite their different
solid fractions (see table \ref{tab:p001}), have
considerably decreased, as shown on Fig.~\ref{fig:hist2636}.
\begin{figure}[!htp]
\flushleft
\includegraphics[height=.95\columnwidth,angle=270]{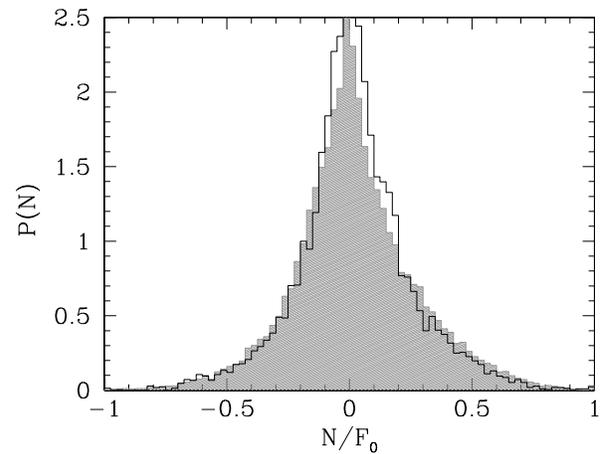}
\caption{Comparisons between probability distribution functions of
normal force values in samples of type A (histogram, in black) and
B (shaded histogram, grey) without RR under
$P^*=0.01$.\label{fig:hist2636}}
\end{figure}
The influence of such differences in the aggregation stage as
those between our samples A and B are therefore expected to fade out
after the systems are compressed to higher pressures and
densities.
\subsection{Packing structure and force patterns\label{subsec:forces}}
The spatial organization of forces in type 2 samples, which we now
discuss, is related to the distribution of force values, and
should determine the ability of given configurations and contact
networks to support stress increments. We first discuss systems
without RR, then with the small RR values we adopted in most cases
(see Table~\ref{tab:param}). We emphasize the role of \emph{force
indeterminacy} and \emph{assembling history} (the collisions by
which cohesive clusters were built) in the final force patterns in
equilibrium under vanishing or low applied stress. Extreme cases
of systems with large RR on the one hand, or without friction on
the other hand, are useful reference situations, which we briefly
examine and discuss. We conclude this part with a discussion of
the main physical implications of the relationships between force
patterns, assembling process, geometry and micromechanical
parameters
\subsubsection{Qualitative aspects of force networks with no RR\label{sec:quali}}
It is instructive to represent the forces carried by the contact
network with a visualization of positive (repulsive) and negative
(attractive) normal forces, as was done on Fig.~\ref{fig:adm1},
showing the force network in one type 1 sample. Figs~\ref{fig:g02}
and \ref{fig:g03}
\begin{figure*}[!htb]
\centering
\includegraphics[width=13.1cm]{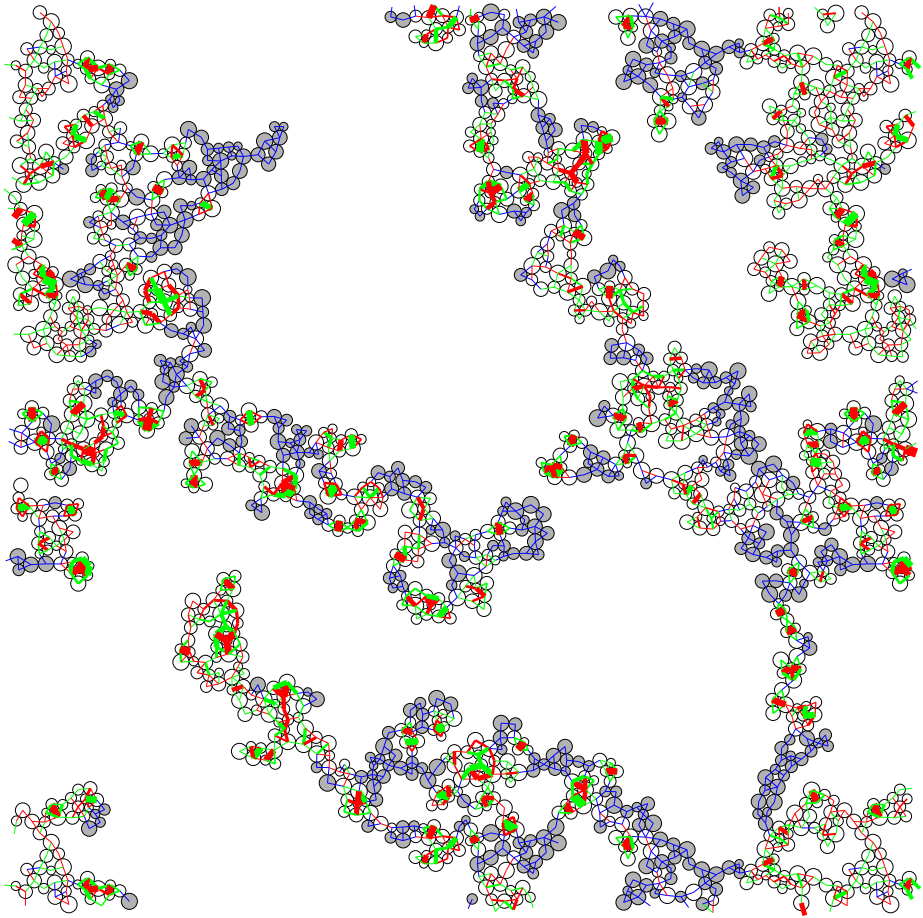}
\caption{\label{fig:g02} (Color online) Sample of type 2 (N=1400),
in equilibrium under $P^*=0$ after aggregation stage, with solid
fraction $\Phi=0.26$ (series B). Same conventions as on
Fig.~\ref{fig:adm1}, except for the blue color corresponding to
contacts carrying a total force below tolerance $10^{-4}F_0$
(deflection $h_0$ and no mobilization of tangential force). Note
the large number of such interactions and the local compensation
of attractions and repulsions in small prestressed clusters.
To help visualize unstressed regions, disks only interacting at contacts bearing
forces below tolerance are filled in light grey.}
\end{figure*}
respectively correspond to equilibrated samples prepared with
method 2 under $P^*=0$ (immediately after the aggregation stage)
and under $P^*=0.01$, without RR. They are represented here with
(approximately) the same scale. Both belong to the ($\Phi_\as=0.26$)
B series. Line widths, which are proportional to the intensity of
the \emph{total} interaction force, \emph{i.e.} to $\norm{{\bf F}}
= \norm{N\nn + T\ta}$, witness the presence, in spite of the low
pressure, of many forces of order $F_0$ (which correspond on the
figures to line thicknesses comparable to particle radii). Stressed
clusters, in loose type 2 samples under $P^*=0$, are separated by
large parts of the interacting network in which contacts carry
vanishing forces: the corresponding normal deflection is $h_0$
(Eqn.~\eqref{eqn:h0}) and there has been no elastic relative
tangential displacement.
\begin{figure*}[!htb]
\centering
\includegraphics[width=10.7cm]{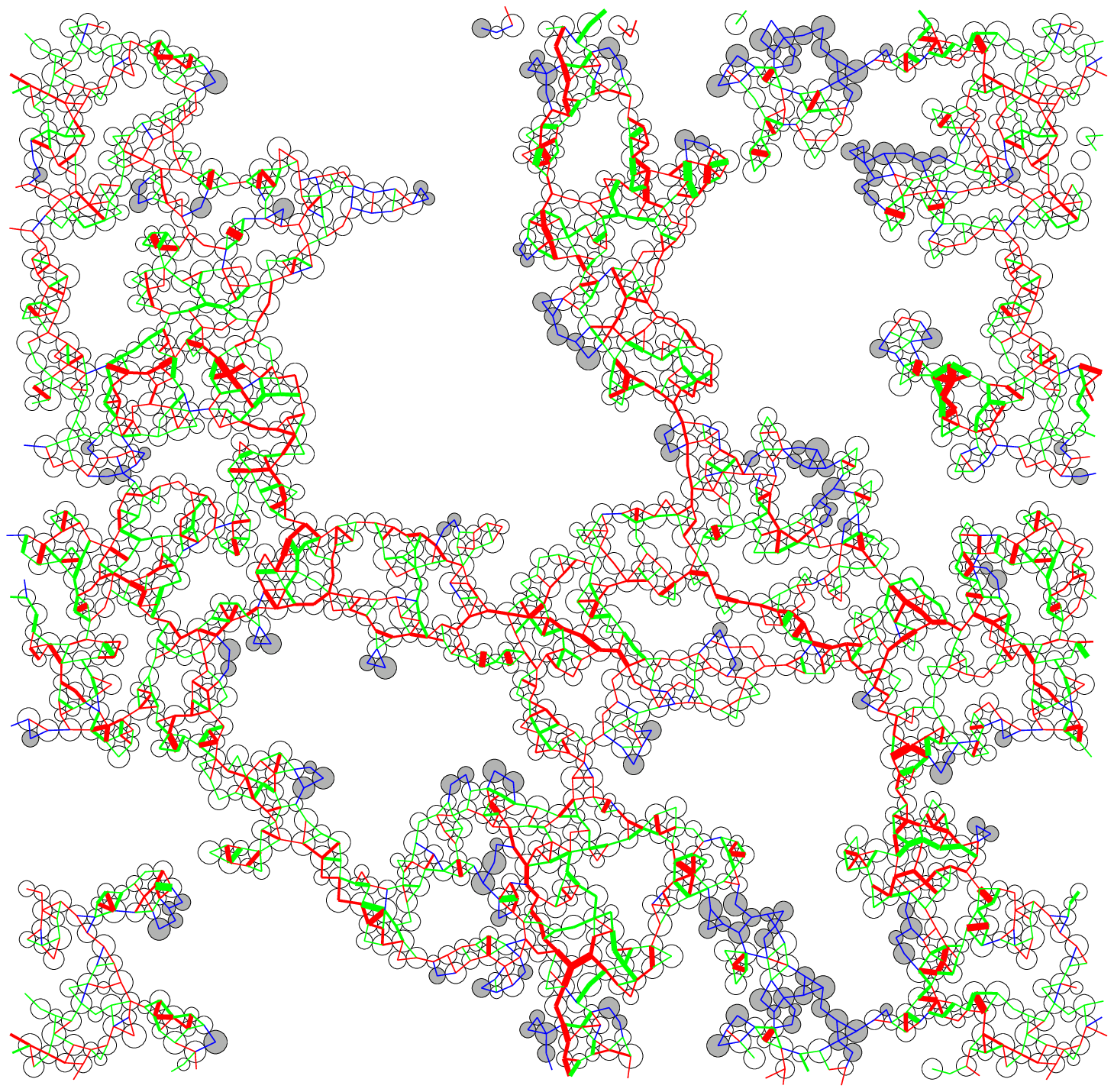}
\caption{\label{fig:g03} (Color online) Sample of
Fig.~\ref{fig:g02}, with same scale and color conventions, in
equilibrium under $P^*=0.01$. The solid fraction increased to
$\Phi=0.39$. The threshold force (used to distinguish blue lines and grey disks) was set to $0.01F_0$.}
\end{figure*}
Attractions (green) and compressions (red) have to compensate for
Eqn.~\eqref{eqn:rumpf2D} to hold true. This compensation appears
to operate on a smaller scale in type 2 samples, because internal
forces were previously balanced within isolated particle clusters.
Such a local balance of forces is quite conspicuous at $P^*=0$
(Fig.~\ref{fig:g02}), in which internal stresses in small clusters
often take the form of a peripheral tension compensating a radial
compression, or the other way round. This contrasts with samples
prepared with method 1 (Fig.~\ref{fig:adm1}), in which the spatial
distribution of forces is more similar to the familiar ``force
chain" pattern of cohesionless systems, although there are of
course compressive and tensile ``chains". Unstressed regions are
rather scarce in type 1 samples, although some areas with smaller
forces are still present. The structure of type 2 samples under
$P^*=0.01$ (Fig.~\ref{fig:g03}) is somewhat intermediate: isolated
stressed clusters are still present, but elongated,
force-chain-like structures emerge.

To characterize such force patterns in a slightly more
quantitative way, one can evaluate a threshold force
$F_{\text{perc}}$ such that contacts carrying a force ${\bf F}$
with $\norm{{\bf F}}\ge F_{\text{perc}}$ percolate through the
sample. Such a criterion was used to identify a ``strong''
subnetwork of force chains in~\cite{RWJM98}. One observes
$F_{\text{perc}}$ of the order of the tolerance $10^{-4}F_0$ in
$\Phi_\as=0.26$ samples with no RR and $P^*=0$, which shows that
stressed regions are isolated ``islands'' within the network.
$F_{\text{perc}}$ raises to slightly less than $0.1F_0$ under
$P^*=0.01$.

Configurations of series A, assembled with $\Phi=0.36$, possess
the same qualitative features, although quantitatively slightly
weaker, due to their higher density. For instance, local stressed
regions are somewhat less isolated, with a threshold force
$F_{\text{perc}}$ between $10^{-3}F_0$ and $8\times 10^{-3}F_0$ at
$P^*=0$.
\subsubsection{Force indeterminacy (without RR)\label{sec:hyperstat}}
The presence of large interaction forces of order $F_0$ in
equilibrated samples is not obviously necessary, and is related to
the assembling process. Let us imagine particles are brought very
slowly, one by one, within interaction range of the previous
network, thus gradually building a unique cluster in equilibrium
in the absence of external stress. One could expect, rather, each
new contact to stabilize with $N=T=0$ and $h=-h_0$. The existence
of non-zero interaction forces in equilibrium is related to the
\emph{hyperstaticity} or \emph{force indeterminacy} of the contact
network. On writing all equilibrium equations for grains and
collective degrees of freedom (\emph{i.e.,} setting acceleration terms
to zero in Eqns.~\ref{eqn:newton}, \ref{eqn:moment} and \ref{eqn:cell}) and
regarding all contact forces as unknowns, the degree of force
indeterminacy $\hhh$ (or degree of hyperstaticity) is the number
of remaining independent unknowns, which cannot be determined by
the equilibrium requirement. If $\hhh=0$, knowing that some
equilibrium forces exist (since an equilibrium state has been
found), then one would necessarily have all interaction forces
equal to zero under $P^*=0$ (since this is one obvious possible
solution). The notion of force indeterminacy has been recently
discussed by different groups in the context of granular
materials, essentially because of the special case of rigid
frictionless grains, for which the contact network is generically
such that forces are uniquely
determined~\cite{OR97b,JNR97b,MO98a,TW99,JNR2000}. The degree of
force indeterminacy is linked to the number of degrees of freedom,
equal to $3N$ (or $3N+2$ if the cell sizes can change), to the
number of contacts $N_c$, the number of distant interactions $N_d$
and the number of independent \emph{mechanisms} or \emph{floppy
modes} $\kkk$ (also called degree of hypostaticity~\cite{JNR2000})
by the following relation (written here for a fixed cell) \be
3N+\hhh = 2N_c+N_d+\kkk \ \mbox{(no RR).} \label{eqn:relhk} \ee A
proof of this simple result (which is classical in structural
engineering), and the relation of numbers $\hhh$ and $\kkk$ to the
\emph{rigidity} and \emph{stiffness} matrices of the contact
network, are recalled in Appendix~\ref{sec:apprig}.
\emph{Mechanisms} are those sets of velocities (or small
displacements, dealt with as infinitesimal) which entail no
relative velocities (or small relative displacements) in contacts.
For distant interactions, only normal relative velocities are
relevant, hence their particular treatment in~\eqref{eqn:relhk}.
In Appendix~\ref{sec:apprig} we  explain how we determine whether
a given configuration is \emph{rigid}, \emph{i.e.}, devoid of
mechanisms (apart from the two global translational motions of the
whole set of grains, rendered possible by the periodic boundary
conditions).
It is customary to relate the level $\hhh$ of force indeterminacy to
the coordination number $z$ in granular materials. However, this
is not possible in general, which motivated our
recalling~\eqref{eqn:relhk} in its complete form.
\eqref{eqn:relhk} can be rewritten, neglecting the very scarce
distant interactions, as
$$
\hhh=N(z-3)+\kkk.
$$
Hence, in the absence of floppy modes, $\hhh=N(z-3)$. However,
there are still a few floppy modes on structures like those of
Fig.~\ref{fig:g02} at the end of the aggregation stage, and this
relation, which predicts a small degree of hyperstaticity relative
to the number of grains (see Tables~\ref{tab:p0} and
\ref{tab:p001}), is only approximate. Some mechanisms are due to
the (exceptional) 1-coordinated disks and others, less trivial,
are associated with larger parts of the structure which are
connected to the rest of the packing \emph{via} one single
2-coordinated disk. This floppiness is obviously related to the
assembling process: before any external pressure is applied,
nothing really requests the aggregates to possess a rigid
backbone. The free motion of mechanisms in assembling method 2 is
largely responsible for the very long equilibration time (see
Fig.~\ref{fig:prep}): such motions entail no restoring force and
no dissipation of kinetic energy. Floppy modes in the final state
obtained with our criteria (Section~\ref{subsec:equilibrium})
being scarce (typically, a few such mechanisms per 1400-disk
sample), we conjecture that they would disappear entirely on
adopting stricter equilibrium requirements in terms of kinetic
energy. If a mechanism survives, it should generically be in
motion with a non-vanishing velocity, as a residual effect of the
initially agitated state.  As the connected aggregate partly folds
onto itself, such motions should eventually create new adhesive
contacts, thereby reducing $\kkk$, until the network becomes
rigid. Once some rigid aggregate is formed in the assembling
process, it will keep the same shape and structure, unless the
collisions and perturbations it subsequently undergoes cause it to
break, because of the limited tensile strength of contacts or
because of the Coulomb inequality. This is the reason why the
initial mean quadratic velocity of isolated grains in method 2
should be compared to $V^*$, as given by~\eqref{eqn:vsep}.

It is easy to see that the closing of one contact can convert an
aggregate from floppy to hyperstatic, the simplest example thereof
being the ``double triangle'' structure of Fig.~\ref{fig:fdt}.
By~\eqref{eqn:relhk}, this small structure, which is rigid ($\kkk=3$
counting the free motions of an isolated object in 2D) has a
degree of force indeterminacy $\hhh=1$.
\begin{figure}[!htb]
\centering
\includegraphics*[height=4cm,angle=270]{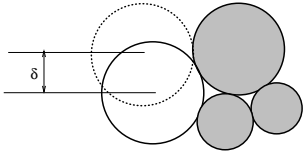}
\caption{\label{fig:fdt} Three (grey) disks initially forming an
isostatic structure, when a fourth one (coming from the left)
adheres to one of them (dotted position), it can roll (this is a
floppy mode) until another (fifth) contact is formed, stabilizing
it in the final position drawn with continuous line. The final
``double triangle'' structure is hyperstatic. Final forces (see
main text) were computed for different initial velocities of the
mobile disk and for different values of impact parameter
$\delta$.}
\end{figure}
This is how the self-stressed clusters of Fig.~\ref{fig:g02} are
formed. Such structures have a strong influence on force values
and force distribution. In particular, we show now that they
entail specific correlations between normal and tangential force
components in contacts.
\subsubsection{Local patterns and specific force orientations}
Fig.~\ref{fig:nt} shows all contact force values as points in the
$N,T$ plane for a 5600 disk sample without RR equilibrated under
$P^*=0$.
\begin{figure}[!htb]
\includegraphics*[width=0.9\columnwidth]{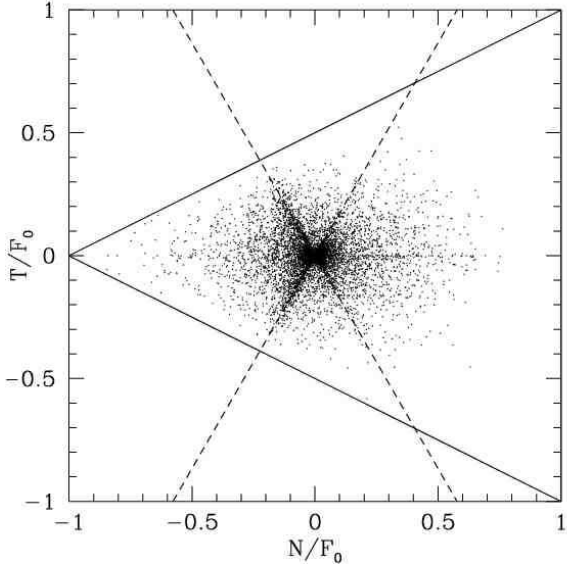}
\caption{\label{fig:nt} Values of normal and tangential contact
forces in a 5600 disk, type 2 sample, in equilibrium under
$P^*=0$, with $\Phi=0.36$ (A configuration). In addition to the
remarkable cross-shaped pattern, marked with dashed lines of
slopes $\pm \sqrt{3}$, note the large number of very small forces,
the numerous points with $\vert T\vert \ll \vert N\vert$ and the
relevance of the value of the friction coefficient ($\mu=0.5$
here), as a small number of forces approach the Coulomb cone.}
\end{figure}
Fig.~\ref{fig:nt} displays a striking X-shaped distribution in the
$N,T$ plane, corresponding to a ratio $T/N$ of $\pm\sqrt{3}$ This
cross pattern fades away in systems which have rearranged to
support $P^*=0.01$, although the corresponding specific $T/N$
ratios are still overrepresented, as shown on
Fig.~\ref{fig:hitheta}.
\begin{figure}[!htb]
\subfigure[$P^*=0$]{
\includegraphics*[width=6cm,angle=270]{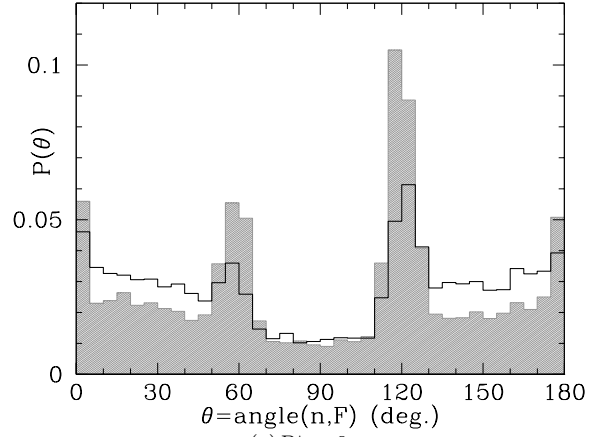}
\label{fig:hitheta1}}
\subfigure[$P^*=0.01$]{
\includegraphics*[width=6cm,angle=270]{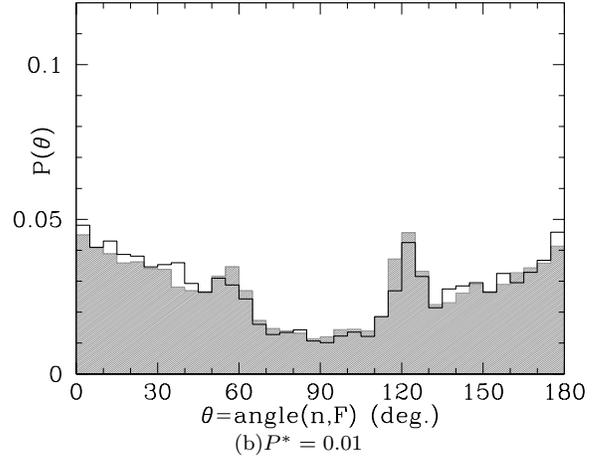}
\label{fig:hitheta2}} \caption{\label{fig:hitheta} Histograms of
angle $\theta$, between normal vector ${\bf n}$ and total contact
force ${\bf F}$. Conventionally, $\theta=0\degr$ for a repulsive
normal force and $T=0$, and $\theta=180\degr$ for a tensile normal
force and $T=0$. Shaded histograms (grey) correspond to B
configurations ($\Phi_\as=0.26$), bold-line non-shaded ones
(black) to A ones ($\Phi_\as=0.36$) }
\end{figure}
The cross pattern of Fig.~\ref{fig:nt} corresponds to angles
$\theta=60\degr$ and $\theta=120\degr$ on Fig.~\ref{fig:hitheta},
and the second graph shows that $\theta=120\degr$ still corresponds
to a peak in the distribution once the sample has been compressed
(and rearranged) to $P^*=0.01$. As other characteristic features
of force patterns in loose type 2 samples, this correlation
between tangential and normal force components is stronger in the
more tenuous networks of series B samples, for which the data are
also represented on Fig.~\ref{fig:hitheta}. The difference between
both sample series tends to disappear on compressing to $P^*=0.01$
(Fig.~\ref{fig:hitheta2}).

The prevalence of ratio $\vert \frac{T}{N}\vert \simeq \sqrt{3}$
is in fact easy to understand. Many disks are in equilibrium with
two contact forces, with two other disks which are themselves
contacting each other, as on Fig.~\ref{fig:f2c}.
\begin{figure}[!hbt]
\includegraphics*[height=3.5cm]{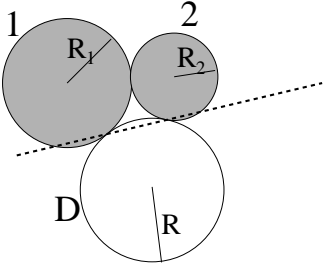}
\caption{\label{fig:f2c}The bottom disk, marked D, of radius $R$,
is in contact with two other disks 1 and 2, themselves touching,
whose radii are $R_1$ and $R_2$. At equilibrium, contact forces on
disk D should be carried by the dotted line joining its 2 contact
points, which determines the ratio of tangential to normal force
components.}
\end{figure}
(Occasionally, a third contact might be present, bearing a much
smaller force, which we neglect in the present argument). In such
a situation, without RR, the three equations expressing the
balance of forces and moments on disk D involve four unknown force
components. Labels corresponding to contacts with disks marked 1
and 2 like on Fig.~\ref{fig:f2c}, one obtains, on counting
positively repulsive normal forces on disk D and tangential forces
with a positive moment:
\be
\ba
N_1 &= N_2\\
T_1 &=-T_2\\
\abs{\frac{T_1}{N_1}}&=\abs{\frac{T_2}{N_2}}=
\sqrt{\frac{R(R_1+R_2+R)}{R_1 R_2}}
\ea
\label{eqn:f2c}
\ee
The ratio in~\eqref{eqn:f2c} varies for the radius distribution we
are using in the present study but its most frequent value,
corresponding to $R_1=R_2=R$, is $\sqrt{3}$. Fig.~\ref{fig:nt0}
shows the same graph as that of Fig.~\ref{fig:nt}, in the case of
a loose packing of disks with the same radii. In agreement with
formula~\eqref{eqn:f2c}, the ``X'' shape is sharply defined.
\begin{figure}[!hbt]
\centering
\includegraphics*[width=0.9\columnwidth]{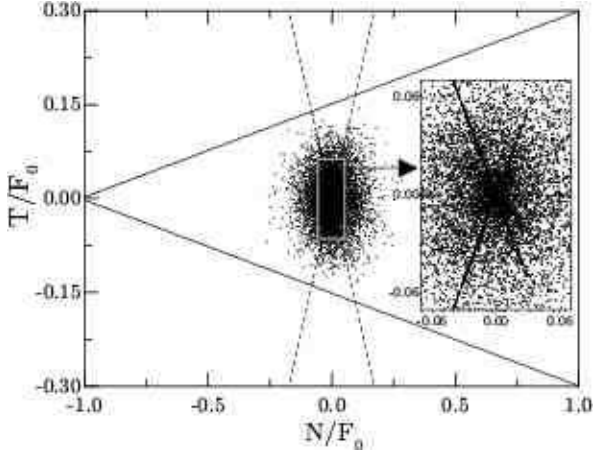}
\caption{\label{fig:nt0} Values of normal and tangential contact
forces in a 5600 disk, type 2 sample of \emph{monodisperse disks}
in equilibrium under $P^*=0$. Note the sharp``X'' shape on
blown-up detail of small forces.}
\end{figure}

To understand the frequency of occurrence of very small $T/\vert N
\vert$ values, let us now consider again the smallest cluster with
force indeterminacy, without RR, which comprises four disks and 5
contacts, as schematized on Fig.~\ref{fig:f2d}. Fig.~\ref{fig:f2d}
shows graphically that the balance of contact forces implies that
the tangential force within the contact corresponding to the
common base of the two triangles should be very small, thereby
explaining the ``dense line'' along the $N$ axis on
Fig.~\ref{fig:nt}. It can be checked by direct inspection that
local simple patterns like those of Figs.~\ref{fig:f2c} and
\ref{fig:f2d} are indeed typical for the forces with ratios $T/N$
around $\pm \sqrt{3}$, or with $\abs{T}\ll \abs{N}$.
\begin{figure}[!htb]
\includegraphics*[width=0.97\columnwidth]{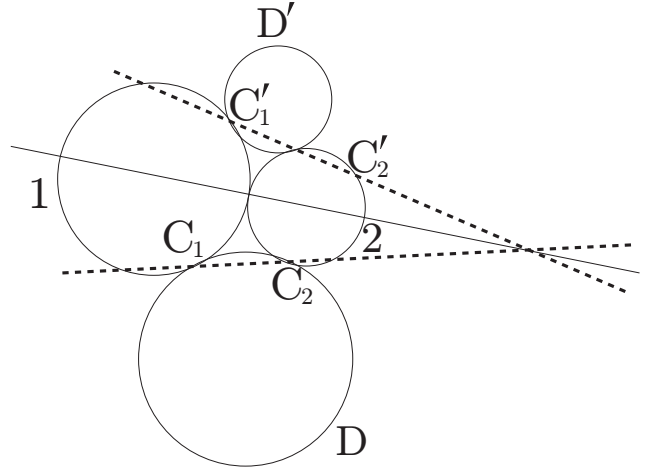}
\caption{\label{fig:f2d} Hyperstatic 4-disk cluster, with 5
contacts. The force at the contact point between 1 and 2 should be
carried by the continuous line joining this point to the
intersection of the dotted lines. Those lines are respectively
defined, as on Fig.~\ref{fig:f2c}, by the two contact points $C_1$
and $C_2$ of lower disk $D$ with disks 1 and 2, and the two
contact points $C_1^{\prime}$ and $C_2{^\prime}$ of upper disk
$D^{\prime}$  with disks 1 and 2. Note that the continuous line is
close to the line of centers, hence a small value of ratio
$\abs{T/N}$ in the contact between 1 and 2.}
\end{figure}
The values of equilibrium forces within such a cluster depend on
how it was built. Without RR three disks forming a triangle
equilibrate with zero contact forces, since there is no force
indeterminacy. On simulating the collision of a fourth disk with
such a triangle (as already sketched on Fig.~\ref{fig:fdt}) all
four particles having the same radius $a/2$, we could observe
final equilibrium situations with contact forces depending on the
impact velocity, provided of course a hyperstatic structure like
that of Fig.~\ref{fig:f2d}, with 5 contacts, was assembled.
Tensile forces equal to $-0.133\times F_0$ in contacts $C_1$ and
$C_2$ of Fig.~\ref{fig:f2d} were created for a contact with an
initial velocity  due to the sole acceleration of the distant
attractive force over distance $D_0=10^{-3}a$ (within a range of
impact parameter $\delta$ defined on Fig.~\ref{fig:fdt}). Larger,
repulsive forces were observed for higher initial approaching
velocities. Self-balanced forces of order $F_0$ therefore
naturally appear in the assembling process.
\subsubsection{Systems with small RR}
With the small level of rolling resistance we have chosen,
$\mu_r=0.005a$ (see Table~\ref{tab:param}), the general features
of the force patterns in systems without RR are only slightly
altered, as apparent on Fig.~\ref{fig:E01},
\begin{figure*}[!ht]
\centering
\includegraphics[width=13.1cm]{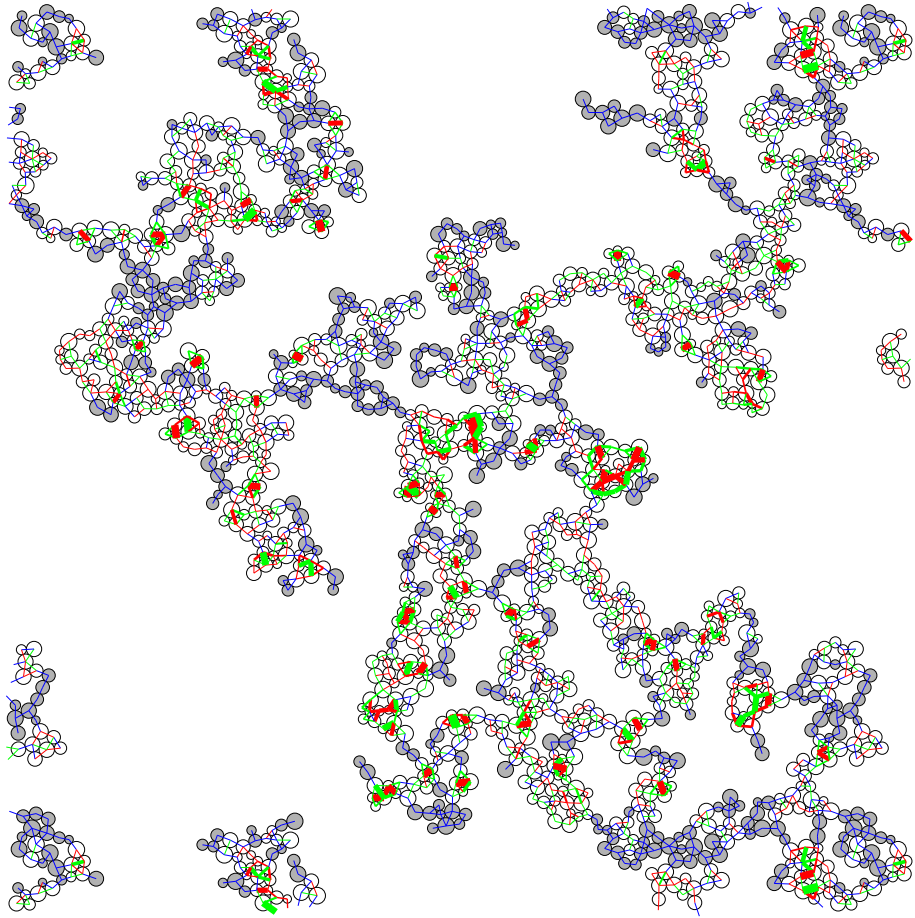}
\caption{\label{fig:E01} (Color online) Same as
Fig.~\ref{fig:g02}, in a (B series) sample with RR, $\Phi=0.26$. Threshold force $10^{-2}F_0$.}
\end{figure*}
which shows the interaction forces in an equilibrated sample of
series B with RR under $P^*=0$. Like in type 2 systems devoid of
RR under $P^*=0$, forces of order $F_0$ only exist in isolated
regions. Note, however, that the small forces outside these
regions with self-balanced stresses do not vanish, but are of
order $\mu_r F_0/a$, a feature which is further commented below.

In principle, the discussion of force indeterminacy and rigidity
is quite different with RR. Contacts now carrying a moment, the
analog of relation~\ref{eqn:relhk} becomes
\be
    3N+\hhh = 3N_c+N_d+\kkk \mbox{ ( RR)} \label{eqn:relhkRR}
\ee
With RR (and as we could check with the method of
Appendix~\ref{sec:apprig}), all connected clusters are rigid. One
may therefore use directly $\kkk=0$ (ignoring the two global
translations) in~\eqref{eqn:relhkRR}. All independent loops
contribute 2 to the degree of force indeterminacy, and the
coordination number corresponding to isostaticity (no floppiness,
no hyperstaticity) is equal to 2. To find self-balanced forces in
a loop, note that each one of the two contacts of any particle in
the loop will carry opposite forces (whence two independent force
components); the resulting torques are then to be compensated by
the rolling moments at the contacts, to be determined with a
number of equations (one per particle in the loop) equal to the
number of unknowns (one per contact in the loop). However, those
moments are severely limited by inequality~\eqref{eqn:ineqRR}. The
constant force ${\bf F}$ transmitted around the loop should then
be of the same order of magnitude as $\mu_r F_0/a$ , hence small.
If we had used the same threshold for blue contacts on
Fig.~\ref{fig:E01} than on Fig.~\ref{fig:g02}, then all contacts
within a loop, because they carry forces above the tolerance level
$10^{-4}F_0$, would have appeared as red (compressive) or green
(tensile). Resetting the threshold to $10^{-2}F_0$, of the order
of $\mu_rF_0/a$, thus enabled us to distinguish the hyperstatic
clusters analogous to the previous case without RR from the new
source of hyperstaticity, the effects of which are limited by the
smallness of the RR parameter $\mu_r$. We could check that the
force threshold $N_{\text{perc}}$ for percolation, as defined
above in paragraph~\ref{sec:quali}, is close to $0.01 F_0$ in that
case. If clusters made with RR, which are (infinitesimally) rigid
could not be broken, no loop should appear because two independent
clusters do not generically collide simultaneously in several
points. The existence of loops in the final structure therefore
witnesses the fragility of tenuous structures which form with a
small level of RR (which is further confirmed by the large scale
changes observed between $P^*=0$ and $P^*=0.01$).

Other features of force distributions and force patterns in
systems without RR, such as the correlations between normal and
tangential contact force components, can still be observed with
the small rolling friction level $\mu_r = 5.10^{-3}a$. The graphs
of Figs.~\ref{fig:nt} and~\ref{fig:hitheta}, if drawn for
configurations prepared in the same way with a small RR, are very
similar. The small RR level used in simulations therefore only
introduces small quantitative differences in that respect, at
least for the parameters of the assembling procedure defined in
Section~\ref{sec:method2}. In the next paragraphs, we investigate,
first, as an instructive limiting case, the effects of large RR,
and then the situations in samples with low RR assembled with
different initial random velocities (as on Fig.~\ref{fig:z-v0}).
\subsubsection{Effect of a large rolling resistance.}
Fig~\ref{fig:RRgrand} shows the analog of Figs.~\ref{fig:g02} and
\ref{fig:E01}, obtained with a large rolling resistance:
$\mu_r=0.5a$ in a 5600 disk sample.
\begin{figure*}[!htb]
\centering
\includegraphics[width=15cm]{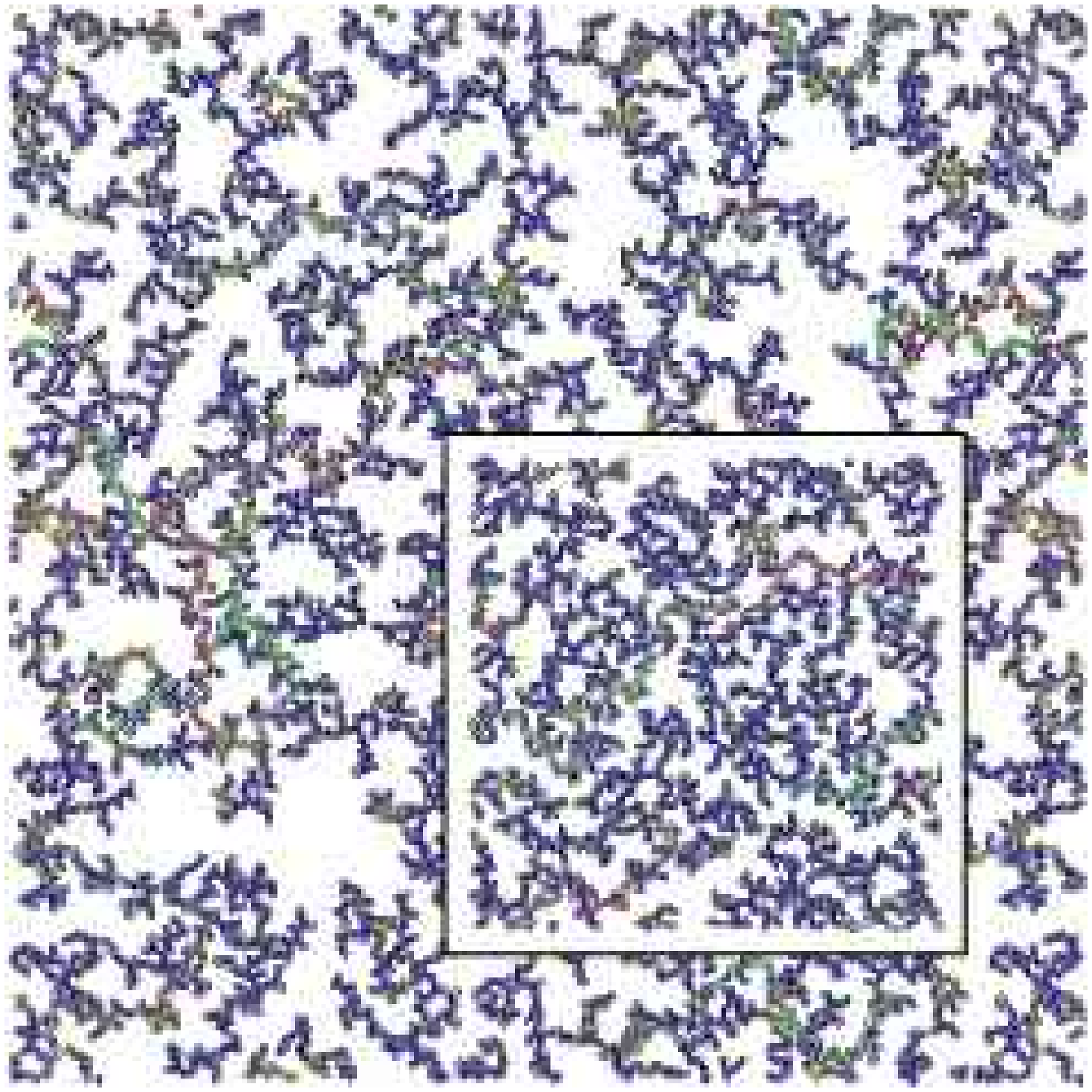}
\caption{\label{fig:RRgrand} (Color online) Same as
Fig.~\ref{fig:E01}, in a sample with large RR, $ \mu_r = 0.5a$,
$N=5600$, and $\Phi=0.26$. Inset: force network in $N=1400$ sample
obtained with low initial mean quadratic velocity $V_0$ and small
RR (corresponding to the bottom left point on
Fig~\ref{fig:z-v0}).}
\end{figure*}
The resulting structure has very few, large loops, hence an
extremely small degree of hyperstaticity, and most contacts carry
but very small forces. The characteristic prestressed clusters of
Figs.~\ref{fig:g02}, \ref{fig:g03} and \ref{fig:E01} have
disappeared. Such packings with large RR therefore approach the
limit in which a simple geometrical rule is adopted to aggregate
particles: in the present case one recovers  the results of the
ballistic aggregation algorithm, stipulating that particles or
clusters move on straight-line trajectories and join to form
larger, rigid objects as soon as they touch. This results in
isostatic, loop-free structures with coordination number 2. The
resulting contact network has no force indeterminacy, and is
consequently not prestressed. Our introduction, in the previous
simulations, of a finite rolling resistance (and a finite friction
coefficient) changes those structures in two respects: first, they
form better connected structures, with loops ; second, they carry
significant self-balanced forces, of the order of the maximum
tensile force in a contact. Those effects are however dependent on
the initial conditions for aggregation, as we now report.
\subsubsection{Effects of initial velocities in aggregation process\label{sec:vinit}}
As shown on Fig.~\ref{fig:z-v0}, the initial mean quadratic
velocity $V_0$ in the aggregation stage of assembling method 2
determines the final coordination number of systems with RR.
Isostatic, loop-free networks are formed with the small RR level
($\mu_r/a=0.005$) used in our systematic simulation series
provided $V_0$ is small enough. The resulting force network, as
displayed as an inset on Fig.~\ref{fig:RRgrand}, approaches a
tree-like, loop-free structure, in which all contact forces vanish
under $P=0$.

Table~\ref{tab:momfv} shows the dependence of coordination numbers
and force values on initial velocity parameter $V_0/V^*$. One
distinguishes three populations of contacts or interactions: those
with repulsive, negative and vanishing normal forces (i.e., below
tolerance level $10^{-4}F_0$), and, likewise, between the average
number of contacts per grains of each kind, respectively
contributing $z_+$, $z_-$ and $z_0$ to the coordination number
$z$. $N_+$ (respectively, $N_-$) is the average value of repulsive
(attractive) normal forces, $N_+^{(2)}$ (resp., $N_-^{(2)}$) the
quadratic average.
\begin{table}[!htb]
\centering
\begin{tabular}{|c|c|c|c|c|c|c|c|c|}\cline{1-9}
$V_0/V^*$  & $z$& $z_+$ &$z_-$& $z_0$ & $ 10^2N_+$ &$N_+^{(2)}$&$ 10^2N_-$ &$N_-^{(2)}$   \\ \hline
$0.095$&2.004&0.12&0.12&1.76&$0.046$&$0.002$&$0.047$&$0.002$\\
$0.95$&2.04&0.38&0.35&1.3&$0.090$&$0.002$&$0.095$&$0.002$\\
$9.5$&2.66&1.17&1.23&0.26&$1.7$&$0.050$&$1.6$&$0.042$\\
$95$&2.96&1.46&1.43&0.07&$5.8$&$0.16$&$5.9$&$0.096$\\
\hline
\end{tabular}
\caption{\label{tab:momfv} Coordination numbers of repulsive,
attractive and unstressed contacts, and values of the
corresponding forces (in units of $F_0$) in samples with RR
prepared at different initial levels of agitation, as on
Fig.~\ref{fig:z-v0}.}
\end{table}
These results illustrate the dependence of the force distribution
on the initial velocity parameter. Force indeterminacy and
significant non-vanishing forces appear as $V_0/V^*$ reaches
values of a few units, with $V_0/V^*=9.5$ corresponding to the
simulation series labelled A and to the force distribution shown
on Fig.~\ref{fig:distrib_forces}.

This set of results therefore bridges the gap between our
\emph{mechanical} studies of cohesive particle aggregation, with
the parameters given in Table~\ref{tab:param} and the preparation
method of Section~\ref{sec:method2}, involving parameter $V_0$,
and the results of geometric algorithms, which are more
traditional in the field of colloid aggregation~\cite{MEA99}.

Geometric changes due to the breaking and rearrangements of
clusters as they aggregate lead to better connected and presumably
less fragile structures, which carry forces of the order of the
maximum tensile force.
\subsubsection{The special case of frictionless disks}
As a complementary study of the opposite extreme case to that of
large RR, we ran some exploratory simulations of
\emph{frictionless, cohesive} grains (also devoid of RR). In the
limit of rigid disks ($\kappa \to \infty$), one knows then that
such assemblies are devoid of force indeterminacy:
$\hhh=0$~\cite{OR97b,MO98a,TW99,JNR2000}. As a consequence, once
large clusters are formed under no external pressure, all contacts
should bear normal forces equal to zero. Such a situation is
depicted on Fig.~\ref{fig:mu0p0}.
\begin{figure}[!htb]
\centering
\includegraphics[width=\columnwidth]{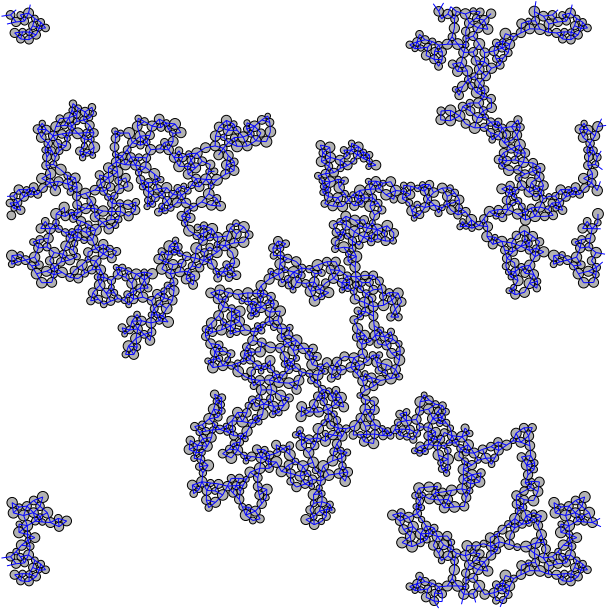}
\caption{\label{fig:mu0p0} (Color online) Same as
Fig.~\ref{fig:g02}, in a sample with $N=1400$, $\Phi=0.26$, no RR
and no friction ($\mu =0$).}
\end{figure}
The aggregate represented on Fig.~\ref{fig:mu0p0} is obviously
floppy. The analog of relation~\eqref{eqn:relhk} is
\be
    2N+\hhh=N_c+N_d+\kkk \mbox{ ($\mu=0$, no RR)} \label{eqn:relhk0}
\ee
(It is customary, on counting degrees of freedom for frictionless disks
or spheres, to discard rotations, which are all irrelevant,
thereby reducing the number of degrees of freedom to $2N$ in the
l.h.s. of~\eqref{eqn:relhk0} ; an alternative is to regard each
rotational degree of freedom as an independent mechanism).

Formula~\eqref{eqn:relhk0} in the frictionless case yields for
$\hhh=0$ a number of floppy modes equal to $2N-N_c-N_d =
(4-z)N/2$. The configuration of fig.~\ref{fig:mu0p0} has a
coordination number $z=3.14$, hence a number of mechanisms larger
than 40\% of the number of particles. Such aggregates are
therefore very floppy, although particles are firmly tied to their
contacting neighbors. Large parts of the particle cluster of
Fig.~\ref{fig:mu0p0} are connected to the rest of the structure by
only one or two contacts, thereby allowing large scale motions
maintaining all contacts. Not surprisingly, the application of a
small pressure $P^*=0.01$ to the system of Fig.~\ref{fig:mu0p0}
produces a very large compression step, resulting in the
configuration shown on Fig.~\ref{fig:mu0p001}.
\begin{figure}[!htb]
\centering
\includegraphics[width=4.5cm]{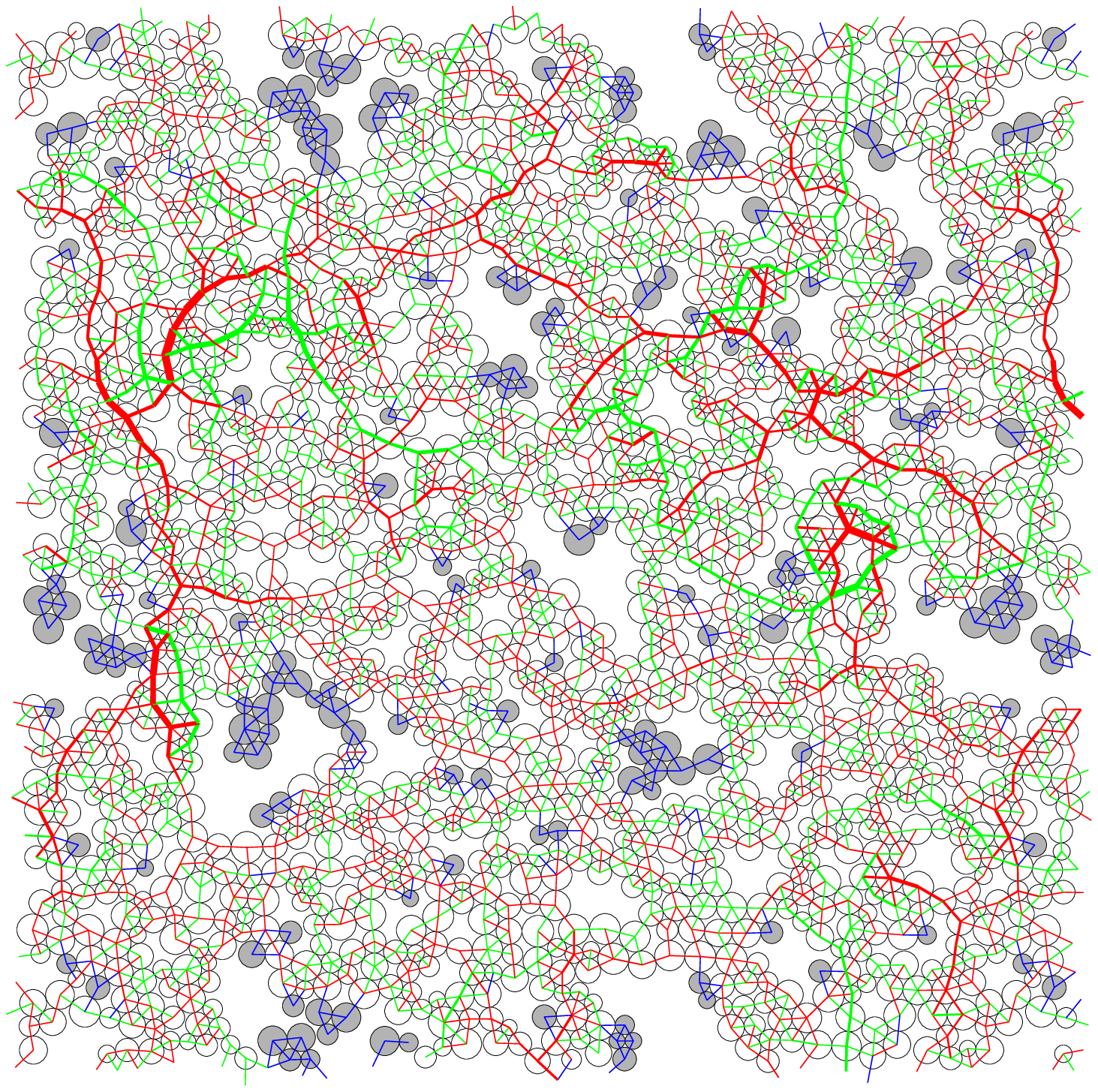}
\caption{\label{fig:mu0p001} (Color online) Sample of
Fig.~\ref{fig:mu0p0} under $P^*=0.01$. $\Phi$ increased to
$0.72$.}
\end{figure}
The coordination number is now $4.01$ (corresponding to a very
small degree of hyperstaticity, due to finite contact stiffnesses
$aK_N/F_0 = 10^4$, as well as to distant interactions), and the
force-carrying network has a rigid subset.

We conclude that assemblies of frictionless, cohesive particles
are rather singular, and do not seem capable of forming stable
loose structures under a non-vanishing confining pressure. It
could of course be conjectured, like in the frictional case, that
floppy networks as shown on Fig.~\ref{fig:mu0p0}, with some
residual motion, would gradually form better coordinated
structures and eventually become rigid, but such an evolution is
too slow to be efficiently followed in our simulations.
\subsection{Discussion}
The study of force values, force distributions and spatial force
patterns we have been presenting here opens quite a few
perspectives that are worth pursuing in more detailed and
quantitative form. In particular, we have left the investigation
of elastic moduli and vibrational eigenmodes of the tenuous
structures formed with method 2 for future work.

However, two qualitative conclusions can be drawn, which might
have broad physical relevance.

First, essentially by direct inspection of force patterns, we
observed that, in loose configurations under relatively low
pressure if compared  to the tensile strength of bonds (as expressed
by $P^*\ll 1$), local arrangements of grains tend to form isolated
self-stressed clusters where forces are of the order of the
maximum tensile force in a contact, $F_0$. Those clusters comprise
any number of grains between a few units to a few tens, keep the
memory of the assembling process, and strongly influence the force
distributions. These features are more apparent at lower
densities. The degree of force indeterminacy $\hhh$ might be a
useful indicator, but is not sufficient in itself, as it is
related to the coordination number, which is very similar in type
1 (dense) and type 2 (loose) systems, and, moreover, does not
account for the role of inequalities~\eqref{eqn:coulomb}
and~\eqref{eqn:ineqRR}. As a general rule, loose cohesive systems
tend to have a wider force distribution when $\hhh$ is larger,
whereas the opposite behavior was observed for confined
cohesionless granular materials~\cite{MJS00}. Dense hyperstatic
clusters in loose packings are connected by regions which bear
very small forces. On increasing the applied pressure by small
amounts, important changes occur, in which these prestressed
regions merge together and large forces tend to organize in
locally preferred directions, as in ``force chains''. Such
structures are likely to play an important role in the mechanics
of loose cohesive granular assemblies. Our mechanical study
stresses the different effects of the two physical origins of
forces -- interparticle attraction and applied pressure -- which
tend to create different geometries, force patterns and force
distributions.

Second, the structure of the loose packings and the forces they
carry are strongly influenced by the assembling conditions. The
relative duration of compression and aggregation processes might produce
results as different as type 1 or type 2 configurations. The velocity of agitation in the initial
assembling stage affects the final coordination number, as shown in Sec.~\ref{sec:vinit}. Such parameters affect
the force patterns as well, and those are also modified if contacts are initially modelled as soft ($\kappa=10^2$), as
in the procedure leading to configurations B.

We expect that mechanical strength properties will also be sensitive to the aggregation process.
\section{Geometric characterization \label{sec:geometry}}
\subsection{Introduction \label{subsec:geometry-intro}}
Aggregation processes are well-known to produce fractal
structures, which have been studied for many years, in particular
with numerical simulations (see ref.~\cite{MEA99} for a review).
Universal fractal regimes due to various types of aggregation
processes (ballistic, diffusion-limited, reaction-limited) are
most conveniently observed in very low density samples. Indeed, an
object of fractal dimension $d_F$ extending over distance $L$ in
$d$ dimensions ($d>d_F$) will have an apparent volume fraction
$\Phi \sim L^{d_F-d}$, which vanishes as $L\to \infty$. Starting
from $N$ isolated particles in a finite volume with periodic
boundary conditions, an aggregation process cannot produce a
fractal geometry over arbitrarily large length scales. In
practice, for low enough values of $\Phi$, the aggregation process
will begin just like in the $\Phi \to 0$ limit, when correlations
between particles can be neglected. Later on, the crowding and
interpenetration of clusters will prevent the fractal behavior to
extend to larger scales
\cite{PFNR04},  
and a classical geometric model to describe this situation is a
dense packing of fractal domains (sometimes called ``blobs'') of
typical diameter $\xi$. $\xi$ is the upper limit of the fractal
regime, and is related to $\Phi$ (see the discussion of eqn. 1
in~\cite{MEA99}) as
\be \label{eqn:blobsize}
    \xi \sim \Phi ^{-\frac{1}{d-d_F}},
\ee
a relation which should be independent of the total sample
diameter $L$, provided $L\gg \xi$. This ``fractal blob'' model is
reminiscent of semi-dilute polymer solutions~\cite{PdG79} and has
been employed in many different physical situations, \emph{e.g.},
silica aerogels~\cite{HAFPJ94}. It has been shown to describe
experimental results on the packing of cohesive
powders~\cite{VQC04,C05}. If such a geometric description applies
to our loose systems, then $\xi$ should be of the order of the
typical size of large density inhomogeneities (dense regions or
holes) in the samples depicted on Fig.~\ref{fig:p0}.

\subsection{Definitions\label{subsec:geometry-definitions}}

Self-similarity is conveniently detected on studying the
\emph{density autocorrelation function} (DACF), as follows. Let
$\chi({\bf r})$ denote the indicator function of solid particles,
taking values $1$ if point ${\bf r}$ is within a solid disk, and
zero otherwise. Then we define the DACF as: \be C(r)
=\ave{\chi({\bf R})\chi({\bf R} + {\bf r})}_{\bf R} =
\frac{1}{A}\int \chi({\bf R})\chi({\bf R} + {\bf r}))d{\bf R},
\label{eqn:cr} \ee with an average over the origin position ${\bf
R}$ over the whole sample surface, of area $A$. On computing
$C(r)$ periodic boundary conditions should be accounted for, so
that position ${\bf R} + {\bf r}$ stays within the simulation
cell. Isotropy ensures that $C(r)$ is only dependent on distance
$r=\norm{{\bf r}}$ in the large sample limit (or on taking its
ensemble average). $C(r)$, by construction, takes the value $\Phi$
(the solid fraction) for $r=0$, and tends to $\Phi^2$ as $r\to
\infty$.

In practice it is convenient to calculate, rather than $C(r)-
\Phi^2$, its Fourier transform, a function of the magnitude $k=\norm{{\bf k}}$
of wavevector ${\bf k}$
by isotropy, which we denote as $I(k)$. $I(k)$ is simply related
to the Fourier transform $\hat \chi$ of the field $\chi({\bf r})$
by:
\be I(k) = \frac{\abs{\hat \chi(k)}^2}{A}. \label{eqn:scatt} \ee
The notation $I(k)$ is of course reminiscent of the scattering
intensity per unit volume for wavevector ${\bf k}$ (as used in
\emph{e.g.} small-angle X-ray or neutron scattering experiments),
which is equal to $I(k)$, up to a ``contrast factor'', replaced by
1 in~\eqref{eqn:scatt}.

A fractal structure with dimension $d_F$ in 2D should have a
power-law decreasing scattering intensity
over some range of $k$: \be I(k) \propto
k^{-d_F} \ \ \ (\frac{2\pi}{\xi}\ll k \ll \frac{2\pi}{a}).
\label{eqn:fractalk} \ee An exponential cut-off of the decreasing
power law behavior of $C(r)$ around $r\sim \xi$ is sometimes
used~\cite{FKZ86,TX88}:
\be \label{eqn:crexp}
    C(r)-\Phi^2= \Phi \left(\frac{r}{\ell}\right)^{d_F - 2}~e^{-r/\xi}
\ee where the length $\ell$, introduced to make $C(r)$ appropriately
dimensionless, is a constant of the order of the average particle
radius. Then the corresponding form of $I(k)$, is given in terms
of Gauss's hypergeometric function
$\Hyp[a,b;c;x]$~\cite{ABRAMOWITZ}:
\be \label{eqn:sk-exp}
 I(k) =\mbox{Cst}+\Phi2\pi\ell^2\Gamma(d_F)\frac{\xi^{d_F}}{\ell^{d_F}}\Hyp\left[\frac{1+d_F}{2},\frac{d_F}{2};1;-\xi^2 k^2\right]
\ee

In 2D, as soon as the particles form one continuous aggregate, the
empty space is split into a set of disconnected holes or
pores. The distribution of sizes and shapes of such holes is
another way to characterize the system geometry.
\subsection{Procedure\label{subsec:geometry-procedure}}
To compute $I(k)$ from the configurations obtained in simulations,
we first discretized the density field $\chi({\bf r})$,
\emph{i.e.,} we considered its values on the points of a regular mesh,
with spacings $\Delta x$ and $\Delta y$ along the edges of the
rectangular cell of the order of $a/100$. $\chi({\bf k})$ was then
evaluated using a two-dimensional FFT algorithm, from which $I(k)$
was deduced by formula~\eqref{eqn:scatt} and orientationally
averaged on binning values of wavevectors ${\bf k}$ according to
$k=\norm{{\bf k}}$.

The field $\chi({\bf r})=0$ defines a set of holes. We
characterize a \emph{hole} labelled as $H$ by the value of its
\emph{equivalent radius} $R_H$. $R_H$ is defined as the radius of
a disk with the same radius of gyration as the hole. Specifically,
if $N_H$ is the number of mesh nodes in the hole, which are
labelled as $i$, $1\le i\le N_H$, and have coordinates $x_i$,
$y_i$, on denoting as ($x_H^c$,$y_H^c$) the coordinates of the
mass center of the hole, one has:
\be \label{eqn:RH}
    R_H = \sqrt{ \frac{2}{N_H} \sum_{i=1}^{N_H}  \left[ (x_i-x_H^c)^2 + (y_i-y_H^c)^2 \right]}
\ee

Holes have complicated shapes and may be
characterized by other quantities such as eccentricity or higher
geometrical moments, but such refinements lie outside the scope of this paper.

For each sample, we record the first weighted moment (or mass average)
of the distribution of hole equivalent radii, $\langle R \rangle_w$,
defined as
\be
    \langle R \rangle_w= \frac{\sum_{H=1}^{\textsl{n}} N_H~R_H}{\sum_{H=1}^{\textsl{n}} N_H}
    \label{eqn:RHave}
\ee
where $\textsl{n}$ is the total number of holes in the sample. In
loose cohesive samples, we obtained a rapid power-law decay for
the shape of this distribution. Definition~\eqref{eqn:RHave},
rather than a simple number average, ensures that the very small
cavities (formed by three or four disks in contact) do not
dominate in the evaluation of the average and $\langle R
\rangle_w$ indeed characterizes the large pores in the loose
packings. However, this definition can only be applied when holes
do not percolate through the aggregate. Thus, we have
restricted the calculation of $\langle R \rangle_w$ to samples
with a non-vanishing confining pressure, $P^* > 0$, in which case
we regard it as an independent measurement of length scale $\xi$.
\subsection{Results}
Functions $I(k)$ are shown on Fig.~\ref{fig:s(k)}, along with
their fits by Eqn.~\ref{eqn:sk-exp}, for $P^*=0$ or $0.01$ with
and without RR, for the configurations of series A
(parameters of Table~\ref{tab:param} and
$\Phi_\as=0.36$). The FFT calculations have been
averaged over different characteristics density maps, and the bars
denote the standard errors. To carry out these fits, we have
applied the Levenberg-Marquardt method for nonlinear least-squares
fittings \cite{M78}.
\begin{figure}[!ht]
\centering \subfigure[Without RR.]{
      \includegraphics*[width=.95\columnwidth]{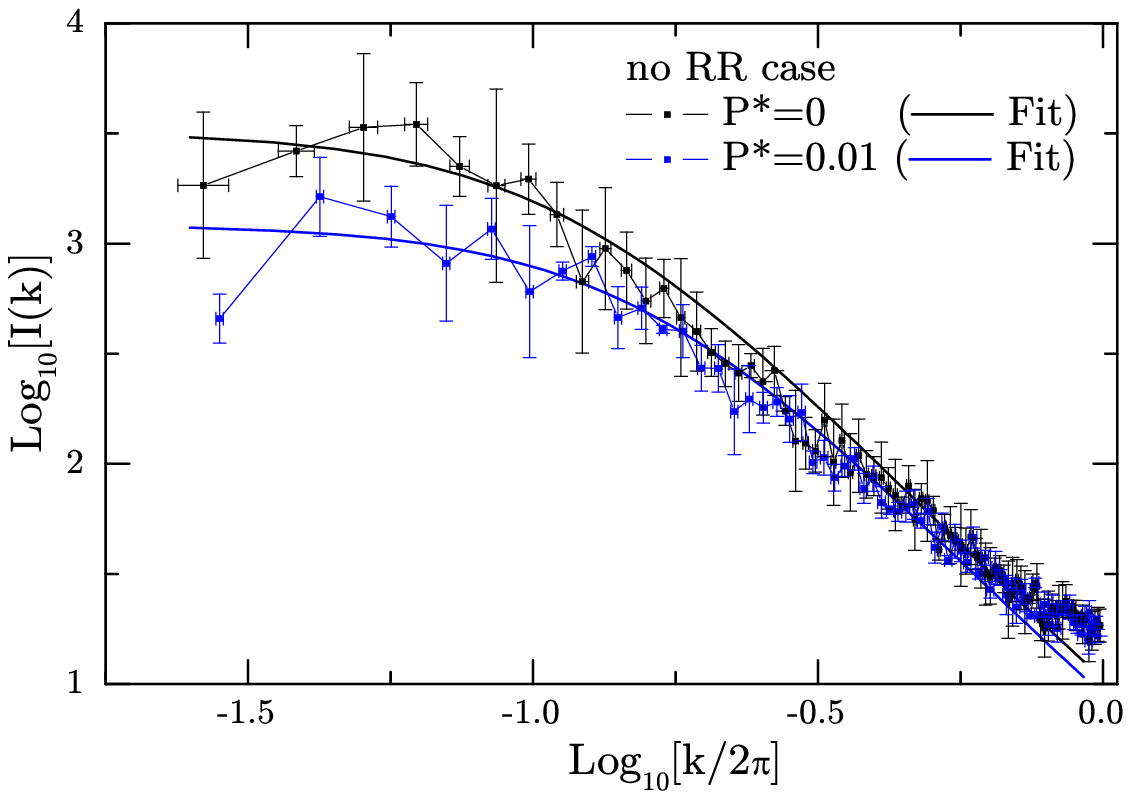}
\label{fig:sk1}} \subfigure[With RR.]{
      \includegraphics*[width=.95\columnwidth]{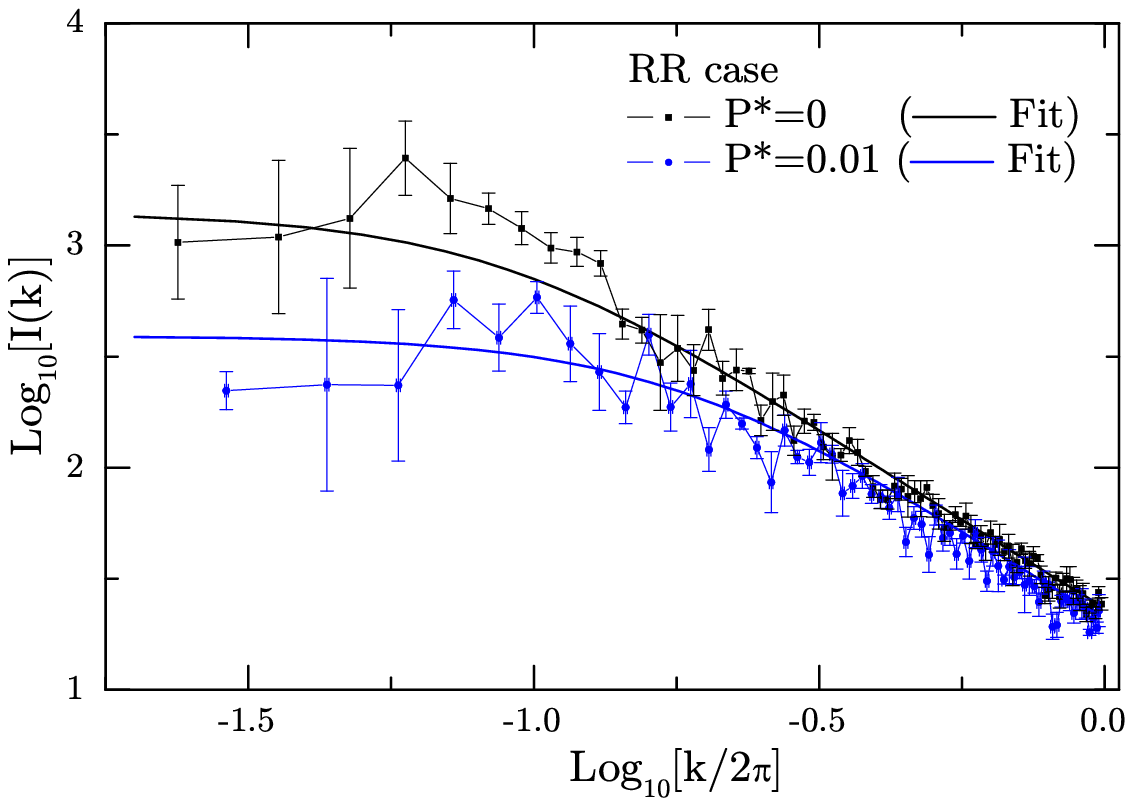}
\label{fig:sk2}} \caption{\label{fig:s(k)} (Color online)
Scattering functions $I(k)$ of samples with and without RR for
$P^*=0$ and $P^*=0.01$, averaged over 4 samples of 1400 disks and
2 with 5600 disks. Fits of data points with Eqn.~\ref{eqn:sk-exp}
are drawn with continuous lines.
Both with and without RR, $I(k)$ is larger for $P^*=0$,
corresponding to larger density fluctuations.
}
\end{figure}
This fitting procedure yields values of $d_F$ and $\xi$ listed in
Table~\ref{tab:fractal}.
\begin{table}[htb]
\begin{tabular}{|c|c|c|c|c|} \hline
\multicolumn{1}{|c|}{} & \multicolumn{2}{|c|}{no RR}   &
\multicolumn{2}{|c|}{RR}         \\ \cline{2-5}
\multicolumn{1}{|c|}{}        & $P^*=0$       & $P^*=0.01$    &
$P^*=0$     & $P^*=0.01$         \\ \hline $d_F$ & $1.925\pm0.024$
& $1.93\pm 0.04$  & $1.53\pm0.04$   & $1.51\pm0.04$ \\ \hline
$\xi/a$                       & $8.29\pm0.15$ & $6.07\pm0.2$ &
$9.3\pm0.4$  & $5.06\pm0.21$
\\ \hline
\end{tabular}
\caption{\label{tab:fractal} Fractal dimension and
fractal blob size obtained on fitting the data of $I(k)$ to
Eqn.~\eqref{eqn:sk-exp}.}
\end{table}
As expected, the fractal dimension is conserved in the compaction
between $P^*=0$ and $0.01$, but the fractal range shrinks. The
marked difference in $d_F$ caused by the introduction of a small
level of RR is remarkable. While self-similar clusters are very
nearly dense ($d_F$ approaching 2) without RR, more open fractal
structures are stabilized on small scales with $\mu_r=0.005a$.
This value of the fractal dimension obtained with RR appears to
coincide, within the error bar, with the value $d_F = 1.55\pm
0.02$ obtained for the ballistic cluster aggregation
model~\cite{MEA84,SMI90}, assuming particles or clusters move on
rectilinear trajectories and stick to one another, forming rigid
objects, as soon as they touch. The tenuous, loop-free structure
of such objects, as previously commented, is retrieved in our
simulations on using a large rolling friction coefficient or a
small level of initial velocity fluctuations. If measured on such
samples as those of Fig.~\ref{fig:RRgrand}, the same result was
obtained, as expected: $d_F = 1.56\pm 0.04$. With small RR or
larger initial velocities, our simulations produce structures
with, apparently the same fractal dimension, but a larger
coordination number.
Another observation from Fig.~\ref{fig:s(k)} is the presence of a
slight bump (maximum) in $I(k)$, for $\frac{2\pi}{k}\simeq 10a$,
(which is not present in the fitted function~\ref{eqn:sk-exp}).
Such a feature is analogous to the peak in the structure factor of
dense particle assemblies, and is likely related to the packing of
aggregates. As the aggregates are requested to be mechanically
rigid they tend to be multiply connected and, at least in 2D,
nearly impenetrable: the maximum in the structure factor is a
signature of steric exclusion.

The evaluation of the first weighted moment $\langle R \rangle_w$
of the distribution of hole equivalent radii, for $P^*=0.01$
yields $ \langle R \rangle_w/a=6.6 \pm 0.2$,
and $\langle R \rangle_w/a=5.47 \pm 0.14$ with RR. As expected,
these results are similar to the values of $\xi$ given in Table~\ref{tab:fractal}.

\section{Conclusions and final remarks \label{sec:final}}

Cohesive packings can form equilibrium structures at very small
solid fractions in qualitative agreements with experiments in fine
and ultrafine powders. Equilibrated configurations are sensitive
to the level of applied pressure, relative to contact tensile
strength $F_0$, as expressed by the dimensionless number $P^*$.
They crucially depend on the assembling procedure, as even a low
pressure ($P^*\ll 1$) can lead to rather compact states if applied
to a initial ``cold" (i.e., with vanishing or low velocities)
granular gas of isolated particles, as in method 1. If, on the
other hand, particles are given some random motion and have time
to stick to one another before having to sustain some stress, as
in method 2, tenuous particle networks and open structures are
obtained. The initial random motion, which is ballistic in our
simulations, could be diffusive in practical situations in which
fine particles are dispersed in a fluid. Some random
\emph{relative} motion of different particles is also present, due
to velocity fluctuations, in situations of global flow or
suspension sedimentation.

Under low pressure, such loose packings carry self-balanced forces
of the order of the maximum tensile force $F_0$ in hyperstatic,
well-connected lumps joined by thinner arms where many contacts
carry vanishing or very small forces. Such structures are
sensitive to the magnitude of initial velocities with which
particles collide on forming aggregates. In general force networks
differ from the usual ``force chain'' patterns of cohesionless
systems, and are associated with different force distributions.
The force balance is strongly influenced by the structure of small
aggregates that are first created on assembling the system. They
evolve very fast as the system rearranges when $P^*$ grows even by
small amounts (from $P^*=0$ to $P^*=0.01$).

Due to the limited strength of contacts with respect to tangential
relative displacement and rolling, force-carrying structures
therefore differ from the ones obtained with geometry-based
algorithms in which any particles or clusters that join form one
unique rigid, unbreakable object. The result of such algorithms is
however retrieved, \emph{in the presence of rolling resistance},
if large strength properties are attributed to contacts (to the RR
parameter $\mu_r$ in particular) or if initial velocities of
colliding grains are kept low enough. In such limits isostatic,
loop-free clusters are formed with coordination number 2.

Micromechanical parameters do otherwise influence the structure of
packings and the initial (self-balanced) forces they carry,
especially those without rolling resistance.

The study of density correlations show that loose configurations
can be regarded as dense packings of self-similar blobs of typical
size $\xi$ (about 10 times as large as the average diameter in our
case), as in fractal clusters produced by colloid aggregation
models. The estimated value of the fractal dimension, with RR, is
compatible with the 2D result for ballistic aggregation, even when
the connectivity (coordination number) is different. We thus
expect different structures of the same density and fractal
dimension to possess, due to the difference in loop numbers and
self-stresses, different mechanical properties.

The fractal dimension appears to be larger in systems without RR.
Thus systems without RR seem to exhibit systematic
\emph{qualitative} peculiarities, and since a small level of
rolling resistance is likely to exist in all realistic models,
this feature should preferably be included in numerical studies.

The effect of a growing pressure, as well as pressure cycles, on
the packing density and internal state will be investigated in a
forthcoming publication~\cite{paper2}. Other immediately related
perspectives are the study of macroscopic tensile and shear
strength in relation to geometric characterizations and
self-balanced forces.

\appendix

\section{Rigidity and stiffness matrices \label{sec:apprig}}
Degrees of force indeterminacy $\hhh$, of velocity indeterminacy
$\kkk$ and their relations are properties of the \emph{rigidity
matrix} $\rig$, which is defined as follows. First, let us denote
as ${\bf U}$ a displacement vector for all degrees of freedom in
the system, \be {\bf U} = \left( ( \tilde {\bf u}_i, \Delta \theta
_i)_{1\le i\le n}, (\epsilon_\alpha)_{1\le \alpha\le 2}\right).
\label{eqn:defu} \ee
 in which one conveniently separates out in
the displacement ${\bf u}_i $ of grain $i$ the part due to the
global strain, thus writing ${\bf u}_i = -\ww{\epsilon}\cdot {\bf
r}_i + \tilde {\bf u}_i$. ${\bf U}$ has dimension $3N+2$ for $N$
disks and 2 strain increments. Then for each one of the $N_c$
contacts, say between $i$ and $j$, the relative displacement of
the contact point (with notations $R_{i,j}$ for the radii, and
$\tij$ for unit tangential vectors as in Sec.~\ref{sec:forces})
\be
\delta {\bf u}_{ij}= \tilde {\bf u}_i +\Delta {\theta}_i\times
R_i \tij -\tilde {\bf u}_j +\Delta {\theta}_j\times R_j \tij +
\ww{\epsilon}\cdot {\bf r}_{ij},
\label{eqn:deprel1}
\ee can be
regarded as providing 2 coordinates to one $2N_c$-dimensional
vector of \emph{relative displacements}, $\delta {\bf u}$.
As~\eqref{eqn:deprel1} expresses a linear dependence of $\delta
{\bf u}$ on ${\bf U} $, one has defined a $2N_c\times(3N+2)$
matrix, which is the \emph{rigidity matrix} $\rig$:
\be
\delta {\bf u}=\rig \cdot {\bf U}
 \label{eqn:deprel}
\ee
All coordinates of ${\bf u}$ and $\delta {\bf u}$ are to be
thought of as small (infinitesimal) increments, for which the
system geometry is fixed. The degree of displacement (or velocity)
indeterminacy $k$ is by definition the dimension of the null space
of $\rig$. The relevant definition of relative displacements
includes all relative motions that are associated to forces or
moments. In the presence of RR, one should include all relative
rotations $\delta\theta_i-\delta\theta_j$ into the components of
$\delta {\bf u}$, the dimension of which thus raises to $3N_c$. On
the other hand, in the absence of friction the tangential relative
displacement of the contact point becomes irrelevant, and $\delta
{\bf u}$ should only include normal relative displacements. In
general all distant attractions between close neighbors should be
dealt with similarly, because only normal forces are transmitted
between such pairs. For future use we just denote as $M$ the
appropriate dimension of the relative displacement vector.

On writing $\delta {\bf u}$ it is most convenient to use a local
basis for each contact, with normal and tangential directions as
coordinate axes. Increments of contact forces, and possibly
moments (with RR), are related \emph{via } the contact law to
$\delta {\bf u}$. Together they define a contact force vector
${\bf f}$, the dimension of which is equal to that of $\delta {\bf
u}$. ${\bf f}$, in a system with RR, also includes rolling
moments at contacts.

Externally applied forces and torques onto the grains, as well as
stresses, define together a vector of external forces $\fext$: \be
\fext = \left(({\bf F}_i, \Gamma_i)_{1\le i\le N},(A
\sigma_{\alpha \alpha})_{1\le\alpha\le 2} \right). \label{eqn:deffext} \ee
$A$ denotes the surface area of the sample, so that the work of
the load for small displacements is just $\fext\cdot {\bf U}$. The
equilibrium relations, stating that contact forces ${\bf f}$
balance the load $\fext$, just read (as one easily checks): \be
\fext = \rigt\cdot {\bf f}, \label{eqn:trig} \ee with the
\emph{transposed} rigidity matrix, $\rigt$. That matrices
appearing in relations~\eqref{eqn:deprel} and \eqref{eqn:trig} are
transposed to each other is just a statement of the \emph{theorem
of virtual work}: the work of external forces in any displacement
vector is $\fext \cdot {\bf U}={\bf f}\cdot\delta {\bf u}$,
provided $\fext$ is related to ${\bf f}$ by~\eqref{eqn:trig} and
$\delta {\bf u}$ is related to ${\bf U}$ by~\eqref{eqn:deprel}. By
definition, the degree of force indeterminacy $\hhh$ is the dimension
of the null space of $\rigt$.

The rank of matrix $\rig$ is $r=N_f-\kkk$, with $N_f$ the number
of degrees of freedom (the dimension of displacement or external
load vectors). This rank $r$ is also the dimension of the range of
the matrix, which is the orthogonal subspace, within the
$M$-dimensional space of relative displacements, to the null space
of its transpose $\rigt$ in the dual space of contact forces.
Hence $r=M-\hhh$. We have obtained
$$
N_f+\hhh=M+\kkk,
$$
which yields, according to the appropriate definition of relevant
relative motions, relations~\eqref{eqn:relhk},
\eqref{eqn:relhkRR}, and \eqref{eqn:relhk0}.

Assuming elastic behavior in the contact (\emph{i.e.}, strict
inequalities in~\eqref{eqn:coulomb} and~\eqref{eqn:ineqRR}, which,
as noted in Section~\ref{subsec:equilibrium},
is the general case at equilibrium), in a quasistatic experiment
contact force increments $\Delta {\bf f}$ relate to relative
displacement increments $\Delta \delta {\bf u}$ with a contact
stiffness matrix $\kk$:
$$
\Delta {\bf f}=\kk\cdot\Delta \delta {\bf u}.
$$
$\kk$ is a square, diagonal matrix, containing coefficients $K_N$,
$K_T$ and (with RR) $K_r$ for each contact. Thus $\kk$ only
contains positive elements, except for the (very scarce) distant
interactions, which contribute the negative normal stiffness
$-F_0/D_0$ in our model. If $\Delta {\bf f}$ balances some load
increment $\Delta \fext$, while $\delta {\bf u}$ corresponds to
the $N_f$-dimensional displacement vector ${\bf U}$, one then has:
$$
\Delta \fext = \sti \cdot {\bf U},
$$
where one has introduced the \emph{stiffness matrix} $\sti$: \be
\sti = \rigt \cdot \kk\cdot\rig. \label{eqn:defstiff} \ee ($\sti$
is traditionally called \emph{dynamical matrix} in the context of
solid-state physics and interactions of atoms or ions in a
crystal~\cite{AM76}). Unlike $\rig$ and $\rigt$, $\sti$ is always
a square, symmetric matrix. It has to be positive definite in
order for the equilibrium state to be stable, because it expresses
the elastic energy associated with small displacements. (In fact,
the full stiffness matrix also contains a small non-symmetric
correction to~\eqref{eqn:defstiff}~\cite{Katalin05} due to the
effect of contact forces prior to the application of the load
increment, which we ignore here.)

By construction, the null space of $\rig$ is contained in the null
space of $\sti$, and coincides with it in the absence of distant
attractions, because $\sti$ is then a positive matrix. In
practice, the positiveness of $\sti$ can be investigated with the
Cholesky algorithm. We applied this method (in a form suitable for
sparse matrices, stored in a ``skyline'' form) to the stiffness
matrix of the contact networks of the simulated equilibrium
configurations. This is how, on finding that $\sti$ was positive
definite, we could conclude that the contact structure was devoid
of mechanisms (or floppy modes, eigenmodes of $\sti$ with
eigenvalue zero) in all cases with $P^*=0.01$. On the contrary,
stiffness matrices associated with contact structures without RR
at $P^*=0$ usually possess some mechanisms, although we argued
that their number $k$ must be small.

{\bf Aknowledgements:} This work has been supported by the
Ministerio de Educaci\'on y Ciencia of the Spanish Government
under contract number BFM2003-1739.\\
J.-N. Roux wishes to thank Dietrich Wolf for
useful contacts and discussions.

\end{document}